\documentclass{aa}  
\usepackage{dcolumn}
\usepackage{aalongtable}
\usepackage{graphicx}
\usepackage{txfonts}
\newcommand{\Ha}{H$\alpha$}
\newcommand{\Hb}{H$\beta$}
\newcommand{\Hg}{H$\gamma$}

\newcommand{\Nl}[3]{#1\,{\sc #2}\,$\lambda{#3}$}
\makeatletter
 \def\hlinewd#1{%
   \noalign{\ifnum0=`}\fi\hrule \@height #1 \futurelet
    \reserved@a\@xhline}
\makeatother
\newcommand{\htopline}{\hlinewd{.8pt}}
\newcommand{\hmidline}{\hlinewd{.2pt}}
\newcommand{\hbotline}{\htopline}
\newcommand{\mcc}[1]{\multicolumn{1}{c}{#1}}

\newcommand{\mcr}[1]{\multicolumn{1}{r}{#1}}
\newcommand{\kms}{km\,s$^{-1}$}
\DeclareMathOperator{\log10}{log10}

\begin{document} 

    \title{Broad-line region structure and kinematics in 
           the radio galaxy 3C\,120
\thanks{Based on observations obtained with the Hobby-Eberly
               Telescope, which is a joint project of the University
               of Texas at Austin, the Pennsylvania State
           University, Stanford University, the Ludwig-Maximilians-Universit\"at
            M\"unchen, and the Georg-August-Universit\"at G\"ottingen.}
}

   \author{W. Kollatschny \inst{1}, 
           K. Ulbrich \inst{1},
           M. Zetzl \inst{1}, 
           S. Kaspi \inst{2,3},
%           \fnmsep
           M. Haas \inst{4}  
          }

   \institute{Institut f\"ur Astrophysik, Universit\"at G\"ottingen,
              Friedrich-Hund Platz 1, D-37077 G\"ottingen, Germany\\
              \email{wkollat@astro.physik.uni-goettingen.de}
            \and
              School of Physics \& Astronomy and the Wise Observatory,
   The Raymond and Beverly Sackler Faculty of Exact Sciences
   Tel-Aviv University, Tel-Aviv 69978, Israel
            \and  
             Physics Department, Technion, Haifa 32000, Israel   
            \and
              Astronomisches Institut, Ruhr-Universit\"at Bochum,
               Universit\"atsstrasse 150, 44801 Bochum, Germany\\
}

%   \date{Received April 12, 2013; accepted July 18, 2013}
   \date{Received 28 March 2014; Accepted 30 April 2014}
   \authorrunning{Kollatschny et al.}
   \titlerunning{Broad-line region in 3C\,120}

% \abstract{}{}{}{}{} 
% 5 {} token are mandatory
 
  \abstract
   {Broad emission lines originate
in the surroundings of supermassive black holes in the centers
 of active galactic nuclei (AGN). These broad-line emitting regions are
 spatially unresolved even for the nearest AGN. The origin and geometry of
broad-line region (BLR) gas and their connection with geometrically
 thin or thick accretion disks is of fundamental importance for the
 understanding of AGN activity.}
%   This line emitting region is spatially not resolved.}
   {One method to investigate the extent, structure, and
 kinematics of the BLR is to study the continuum and line profile variability
 in AGN. We selected the radio-loud Seyfert 1 galaxy 3C120 as a target for 
 this study.}
   {We took spectra with a high signal-to-noise ratio 
 of 3C\,120 with the 9.2m Hobby-Eberly Telescope 
between Sept.\ 2008 and March 2009. 
In parallel, we photometrically monitored the continuum flux
 at the Wise observatory.
We analyzed the continuum and line profile variations in detail
(1D and 2D reverberation mapping) and modeled the geometry
 of the line-emitting regions based on the line profiles.}  
% Furthermore we model the rotational and turbulent velocities in the
%   line-emitting region based on observed full-width at half maximum
%   line values and $\sigma_{\mathrm{line}}$ of the variable
%    broad emission lines.}
   {
We show that the BLR in 3C\,120 is stratified
with respect to the distance of the line-emitting regions
from the center with respect to the line widths (FWHM)
of the rms profiles and with respect to the variability amplitude of the
emission lines. The emission line wings of H$\alpha$ and H$\beta$
respond much faster than their central region. This is explained
by accretion disk models. In addition,
these lines show a stronger response in the
red wings. However, the velocity-delay maps of the helium lines
% \ion{He}{i}\,$\lambda 5876$ and \ion{He}{ii}\,$\lambda 4686$ lines
 show a stronger response in the blue wing.
Furthermore, the \ion{He}{ii}\,$\lambda 4686$ line responds
faster in the blue wing in contradiction to
observations made one and a half years later when the galaxy was
in a lower state.
The faster response in the blue wing is an indication
for central outflow motions when this
galaxy was in a bright state during our observations.
The vertical BLR structure in 3C~120 coincides with that
of other AGN. We confirm the general trend: 
 the emission lines of narrow line AGN originate at larger
distances from the midplane than AGN with broader
emission lines.} 
%   {The derived geometries bla bla}
{}

   \maketitle
%
%________________________________________________________________

\section{Introduction}

The variable radio source 3C\,120 has been identified
to be a distant Seyfert 1 galaxy
of redshift 0.0334 by
Burbidge\cite{burbidge67} as early as 1967.
Later on, French \& Miller\cite{french80} and Oke et al.\cite{oke80}
observed short-term variations (i.e. within of one year)
and long-term spectral variations
in the continuum and in the broad emission lines during
an observing period from 1967 to 1980.
Peterson et al.\cite{peterson98} carried out
a spectral variability campaign
of 3C\,120 during a period of eight years from 1989 to 1996. 
They derived a delay  of
 $\tau = 44.^{+28.}_{-20.}$ days
of the integrated H$\beta$ emission line with respect to the
variable continuum flux.
The value of this delay - that is the distance of the line-emitting
 region from the central ionizing source -
had a large error because
the continuum and emission-line light curves were not sampled
densely. Their mean and median sampling rate was 50 and 11 days, respectively.

We carried out an additional spectral variability campaign of 3C\,120
with the 9.2m Hobby-Eberly Telescope (HET) in the years 2008 and 2009
to study in detail variations in the
integrated line fluxes and in the profiles of the optical Balmer and
helium lines.
The study of variations in
emission-line profiles of active galactic nuclei 
contains information about
the structure and kinematics of the central line-emitting regions in Seyfert 1
galaxies in combination with model calculations.
%Different delays of different emission line segments have been detected
%in some cases (e.g. Gaskell\citealt{gaskell88}).
Relative variations in individual segments of emission lines
with respect to each other
were verified before, for instance, in variability campaigns of the
UV \ion{C}{iv}\,$\lambda 1550$ line
in NGC\,4151, NGC\,5548 (Gaskell \citealt{gaskell88}, Korista et
 al.\citealt{korista95})
or in the Balmer lines
of NGC\,5548, NGC\,4593 (Kollatschny \& Dietrich\citealt{kollatschny96},
Kollatschny \& Dietrich\citealt{kollatschny97}).
There have been indications of a shorter delay in
the red line wings than in the
blue wings in all these variability campaigns.
Detailed two-dimensional
(2D) reverberation-mapping studies have been carried out so far
only for a few galaxies
(Kollatschny\citealt{kollatschny03};  Bentz et al.\citealt{bentz10};
Grier et al.\citealt{grier13}).
Grier et al.\cite{grier13} monitored three spectral lines of 3C\,120
in the year 2010 when this galaxy was in a low state. We compare
their findings with 
the results of our variability campaign. 

Line profile studies confirmed the general picture
 that the broad-line emitting
region (BLR) is gravitationally bound and that the emission lines 
originate in flattened accretion disk structures with additional indications
of inflow or outflow motions.
The analyses of the integrated line-intensity variations
and their line profile variations are important tools for studying
the central BLR in active galactic nuclei (AGN).
In addition, the line profiles - that is, their 
observed full-width at half maximum line values
   (FWHM) and $\sigma_{\mathrm{line}}$ values -
contain information on the rotational and turbulent velocities
 in the line-emitting regions above the accretion disk
(Kollatschny \& Zetzl\citealt{kollatschny11,kollatschny13a,kollatschny13b,kollatschny13c}). Based on these velocity studies, we were able to determine
the heights of the line-emitting regions
above the midplane
in combination with the line-emitting distances from the ionizing
central source for a few active galaxies.
Here we present additional information regarding the vertical BLR structure
in 3C\,120 by modeling their line profiles, as we have done before for four
other Seyfert galaxies
(Kollatschny \& Zetzl\citealt{kollatschny13b,kollatschny13c}),
and we compare the BLR structures with each other.

The paper is arranged in the following way:
in Section 2 we describe the observations taken with the HET
Telescope. 
In Section 3 we present our data analysis
and results on the structure and kinematics of the 
central BLR in 3C\,120.
In Section 4 we discuss the results of our variability campaign
compared with other campaigns of this galaxy.
Finally, we analyze the BLR structure in this galaxy
and compare it with that of other Seyfert galaxies. 
A short summary is given in Section~5.

\section{Observations and data reduction}

 We took optical spectra of the AGN in the Seyfert galaxy 3C\,120
 with the HET telescope at McDonald Observatory 
 at 31 epochs between September 17, 2008, and March 16, 2009.
 The log of our spectroscopic observations is given in Table 1.
\begin{table}
\tabcolsep+6mm
\caption{Log of spectroscopic observations of 3C\,120 with HET}
\centering
%\vspace{3mm}
\begin{tabular}{ccr}
\hline 
\noalign{\smallskip}
Julian Date & UT Date & Exp. time \\
2\,400\,000+&         &  [sec.]   \\
%(1) & (2) & (3) & (4) \\ %& (5) & (6) & (7) & (8) & (9) \\ 
%\noalign{\smallskip}
\hline 
%\noalign{\smallskip}
54726.907 & 2008-09-17 & 1200.0 \\
54729.897 & 2008-09-20 & 1200.0 \\
54739.873 & 2008-09-30 &  600.0 \\
54740.870 & 2008-10-01 &  600.0 \\
54747.857 & 2008-10-08 &  600.0 \\
54748.847 & 2008-10-09 &  600.0 \\
54749.850 & 2008-10-10 &  600.0 \\
54752.834 & 2008-10-13 &  600.0 \\
54759.967 & 2008-10-20 &  600.0 \\
54762.815 & 2008-10-23 &  600.0 \\
54771.774 & 2008-11-01 &  600.0 \\
54775.771 & 2008-11-05 &  600.0 \\
54776.787 & 2008-11-06 &  600.0 \\
54779.773 & 2008-11-09 &  600.0 \\
54783.905 & 2008-11-13 &  600.0 \\
54787.745 & 2008-11-17 &  600.0 \\
54789.737 & 2008-11-19 & 1200.0 \\
54790.736 & 2008-11-20 &  600.0 \\
54793.723 & 2008-11-23 & 1200.0 \\
54807.681 & 2008-12-07 &  600.0 \\
54808.828 & 2008-12-08 &  600.0 \\
54822.652 & 2008-12-22 &  600.0 \\
54830.622 & 2008-12-30 &  600.0 \\
54841.600 & 2009-01-10 &  600.0 \\
54845.725 & 2009-01-14 &  600.0 \\
54866.691 & 2009-02-04 &  600.0 \\
54876.637 & 2009-02-14 &  600.0 \\
54881.635 & 2009-02-19 &  600.0 \\
54887.617 & 2009-02-25 &  600.0 \\
54893.604 & 2009-03-03 &  600.0 \\
54906.576 & 2009-03-16 &  600.0 \\ 
%\noalign{\smallskip}
\hline 
\vspace{-.7cm}
\end{tabular}
\end{table}

\begin{table}
\tabcolsep+8mm
\caption{Log of photometric observations with the Wise and Cerro Armazones
 telescopes}
\centering
\vspace{-0.5mm}
\begin{tabular}{ccc}
\hline 
\noalign{\smallskip}
Julian Date & UT Date & Filter \\
2\,400\,000+&         &      \\
%(1) & (2) & (3) & (4) \\ %& (5) & (6) & (7) & (8) & (9) \\ 
%\noalign{\smallskip}
\hline 
%\noalign{\smallskip}
54745.576 & 2008-10-06 & V,R\\
54745.584 & 2008-10-06 & V,R\\             
54788.625 & 2008-11-18 & I' \\     
54790.625 & 2008-11-20 & I' \\
54791.625 & 2008-11-21 & I' \\
54792.625 & 2008-11-22 & I' \\
54793.625 & 2008-11-23 & I' \\
54794.625 & 2008-11-24 & I' \\
54796.625 & 2008-11-26 & I' \\
54797.304 & 2008-11-26 & V,R\\
54797.625 & 2008-11-27 & I' \\
54798.625 & 2008-11-28 & I' \\
54799.625 & 2008-11-29 & I' \\
54800.625 & 2008-11-30 & I' \\
54802.625 & 2008-12-02 & I' \\
54803.297 & 2008-12-02 & V,R\\
54803.305 & 2008-12-02 & V,R\\
54803.625 & 2008-12-03 & I' \\
54804.304 & 2008-12-03 & V,R\\
54804.312 & 2008-12-03 & V,R\\
54804.322 & 2008-12-03 & V,R\\
54804.331 & 2008-12-03 & V,R\\
54804.625 & 2008-12-04 & I' \\
54805.296 & 2008-12-04 & V,R\\
54805.303 & 2008-12-04 & V,R\\
54805.331 & 2008-12-04 & V,R\\
54808.342 & 2008-12-04 & V,R\\
54808.350 & 2008-12-04 & V,R\\
54808.357 & 2008-12-04 & V,R\\
54811.327 & 2008-12-10 & V,R\\
54817.282 & 2008-12-16 & V,R\\
54817.508 & 2008-12-16 & V,R\\
54818.289 & 2008-12-17 & V,R\\
54819.297 & 2008-12-18 & V,R\\
54825.353 & 2008-12-24 & V,R\\
54826.432 & 2008-12-25 & V,R\\
54837.205 & 2009-01-05 & V,R\\
54837.213 & 2009-01-05 & V,R\\
54838.203 & 2009-01-06 & V,R\\
54839.332 & 2009-01-07 & V,R\\
54841.600 & 2009-01-10 & V,R\\    
54852.172 & 2009-01-20 & V,R\\ 
54853.170 & 2009-01-21 & V,R\\
54856.201 & 2009-01-24 & V,R\\
54859.178 & 2009-01-27 & V,R\\
54866.192 & 2009-02-03 & V,R\\
54867.180 & 2009-02-04 & V,R\\
54893.604 & 2009-03-03 & V,R\\     
54894.191 & 2009-03-03 & V,R\\
54895.192 & 2009-03-04 & V,R\\
54903.304 & 2009-03-12 & V,R\\
54905.205 & 2009-03-14 & V,R\\
54908.238 & 2009-03-17 & V,R\\
54920.215 & 2009-03-29 & V,R\\
55064.527 & 2009-08-21 & V,R\\
55064.535 & 2009-08-21 & V,R\\
55098.571 & 2009-09-24 & V,R\\
55117.533 & 2009-10-13 & V,R\\
%\noalign{\smallskip}
\hline 
\end{tabular}
\end{table}
The obtained spectra span a period of 179.7 days.
The median interval between the individual observations
was 4.1 days and the average interval was 5.8.
During the first two months of our campaign we took 19 spectra with an average
interval of 3.5 days.
In some cases we acquired spectra at intervals of only one day.

All spectroscopic observations were performed under identical instrumental
conditions with the
Marcario Low Resolution Spectrograph (LRS)
mounted at the prime focus of HET. The detector was
a $3072\times1024$ 15 $\mu$m pixel Ford Aerospace CCD with 2x2 binning. 
The spectra cover the wavelength range from 4200\,\AA\
to 6900~\AA\ (LRS grism 2 configuration)
 in the rest frame of the galaxy
with a resolving power of 650 at 5000\,\AA\ (7.7\,\AA\ FWHM).
All observations were taken with exposure times of 
10 to 20 minutes, which in most cases yielded a     
S/N  of at least 100
% $>$  100
per pixel in the continuum.
The slit width was fixed to
2\arcsec\hspace*{-1ex}.\hspace*{0.3ex}0 projected on the
sky at an optimized position angle to minimize differential refraction.
Furthermore, all observations were taken at the same airmass
thanks to the particular design feature of the HET.
We extracted seven columns from each of our object spectra, 
corresponding to 3\arcsec\hspace*{-1ex}.\hspace*{0.3ex}3. The spatial
resolution was 0\arcsec\hspace*{-1ex}.\hspace*{0.3ex}472 per binned pixel.

Both HgCdZn and Ne spectra were taken after each object exposure
to enable the
wavelength calibration. Spectra of different standard stars were
observed for flux calibration as well.

The reduction of the spectra (bias subtraction, cosmic ray correction,
flat-field correction, 2D-wavelength calibration, night-sky subtraction, and
flux calibration) was made in a homogeneous way with IRAF reduction
packages (Kollatschny et al., \citealt{kollatschny01}). 
The spectra were not corrected for the variable atmospheric absorption
in the B band.

 Great care was taken to ensure high-quality intensity and wavelength
calibrations to keep the intrinsic measurement errors very low
(Kollatschny et al., \citealt{kollatschny01,kollatschny03,kollatschny10}).
%For a discussion of the intrinsic measurement error, we refer to 
% Gaskell \& Peterson (\cite{gaskell87}). 
Our galaxy spectra as well as our
 calibration star spectra were not always taken
 under photometric conditions. 
Therefore,
all spectra were calibrated to the same absolute
[\ion{O}{iii}]\,$\lambda$5007 flux of
$3.02 \times 10^{-13} \rm erg\,s^{-1}\,cm^{-2}$
(taken from Peterson et al.\citealt{peterson98}).
The flux of the narrow emission line [\ion{O}{iii}]\,$\lambda$5007
is considered to be constant on time scales ranging from one to a few years
(Peterson et al.\citealt{peterson13}).

% The spatially unresolved structure of the narrow-line region in Mrk\,926
% has been verified before  for utilizing this internal calibration method. 
The accuracy of the [\ion{O}{iii}]\,$\lambda$5007 flux calibration
 was tested for all forbidden emission lines in the spectra.
 We calculated difference spectra for all epochs
 with respect to the mean spectrum of our variability campaign.
Corrections for both small spectral shifts ($<$ 0.5 \AA )
 and small scaling factors were executed
 by minimizing the residuals of the narrow emission lines in the
 difference spectra. 
 All wavelengths were converted to the rest frame of the galaxy (z=0.03302).  
Throughout this paper, we assume that H$_0$~=~70~km s$^{-1}$ Mpc$^{-1}$.
A relative flux  accuracy on the order of 1\% was achieved for most of
 our spectra.

In addition to our spectra of 3C120 taken with the HET (H)
we obtained photometric data taken with the 1m telescope
at the Wise observatory (W) of the Tel-Aviv University in Israel.
The 1m telescope is equipped with the 1340$\times$1300 pixels PI-CCD camera
which has a 13'$\times$13' field of view with a scale of 0.58 arcsec/pix.
Observations were carried out in Bessel V and R bands with exposure
times of 5 minutes.
The images were reduced in the standard way using IRAF
routines. Broad-band light curves were produced
by comparing their instrumental magnitudes with those of non-variable
stars in the field (see, e.g., Netzer et al.\citealt{netzer96}, for
more details). The quoted uncertainties on the photometric
measurements include the fluctuations due to photon statistics
and the scatter in the measurement of the non-variable stars used.

During November and December 2008,
% $i$
I' band measurements were taken with the
40cm monitoring telescope of the Universit\"atssternwarte
Bochum near Cerro Armazones in Chile (Ramolla et al.,\citealt{ramolla13}).
The filter I' is the i-band of PANSTARRS, similar to SLOAN $i${}.
This filter is centered on 7700 \AA{} with a width of 1500 \AA{}.
Per night, ten dithered 60\,s exposures with a size of
27$\arcmin$ $\times$ 41$\arcmin$
were reduced in a standard manner and then combined.
Light curves were extracted using 15$\arcsec$
apertures and five non-variable
stars on the same images and of similar brightness as 3C120.
A list of the photometric observations is given in Table 2.
%\vspace{-3mm}
\section{Results and discussion}
\subsection{Continuum and spectral line variations}
\begin{figure*}
\centering
% \hbox{
%\includegraphics[bb=40 90 380 700,width=9.12cm,angle=270]{2.21_all_spectra.eps}
%\includegraphics[width=5cm,angle=0 ]{2.21_all_spectra.eps}
%\includegraphics[width=15cm,angle=0 ]{Fig1_all_spectra_neu.eps}
%\includegraphics[width=15cm,angle=0 ]{Fig1_all_spectra_wo_shift.eps}
%\includegraphics[width=10cm,angle=0 ]{fig1_all_spectra.eps}
%\includegraphics[width=10cm,angle=0 ]{fig1_all_spectra_wo_shift.eps}
%\includegraphics[width=14cm,angle=0 ]{fig1_all_spectra_wo_shift_2.eps}
\includegraphics[width=13.4cm,angle=0 ]{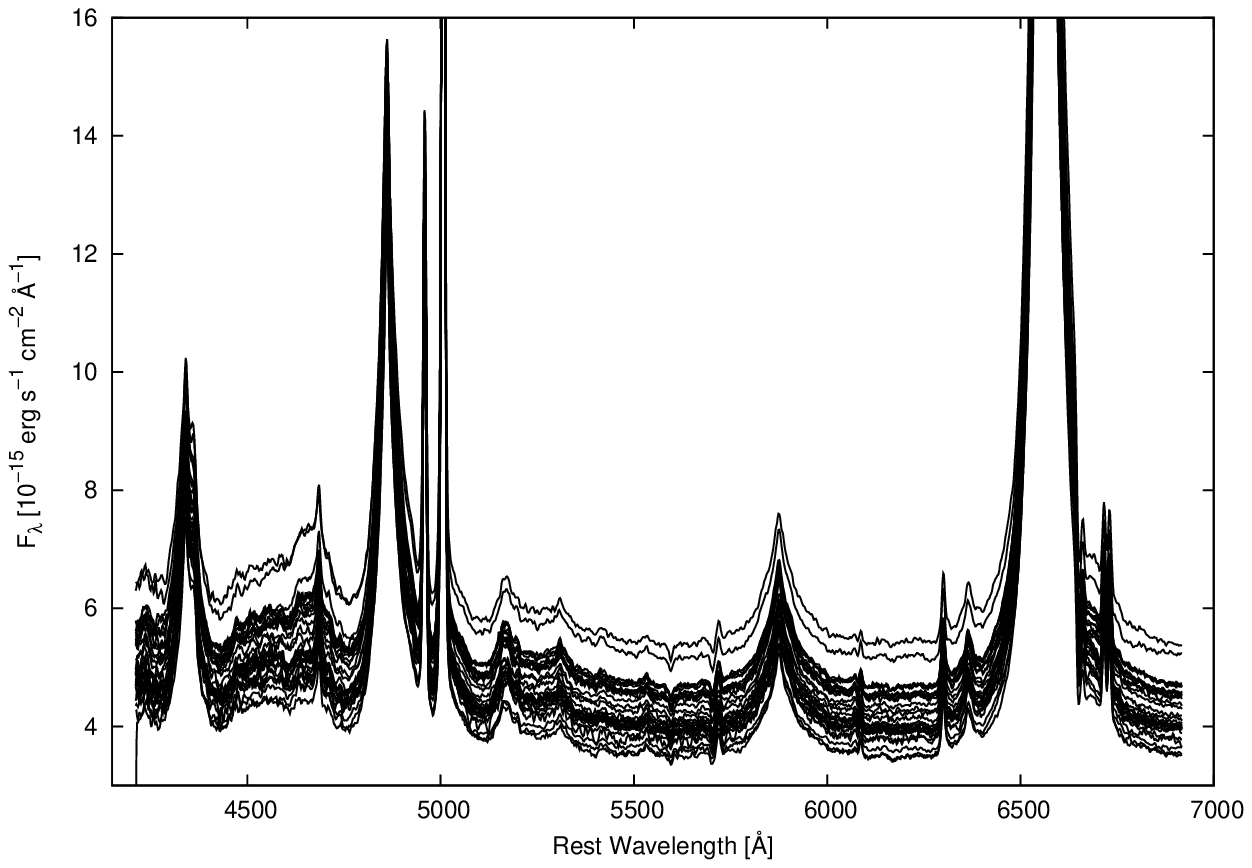}
%}
       \vspace*{-2mm} 
  \caption{Optical spectra of 3C\,120 taken with the HET telescope}
% showing continuum and \ion{He}{ii}\,$\lambda 4686$ line variations.}
   \label{2.21_all_spectra.eps}
%\end{figure*}
%
%----------------------------------------------------------------------------- 
%                                                
%   \begin{figure*}
%    \includegraphics[bb=40 90 380 700,width=9.12cm,angle=-90]{2.25_avg_rms.eps}
%    \includegraphics[width=10cm,angle=0]{2.25_avg_rms.eps}
%    \includegraphics[width=15cm,angle=0]{Fig2_avg_rms_ruhesys_label.eps}
%    \includegraphics[width=15cm,angle=0]{fig2_avg_rms.eps}
%    \includegraphics[width=14cm,angle=0]{fig2_avg_rms.eps}
    \includegraphics[width=13.4cm,angle=0]{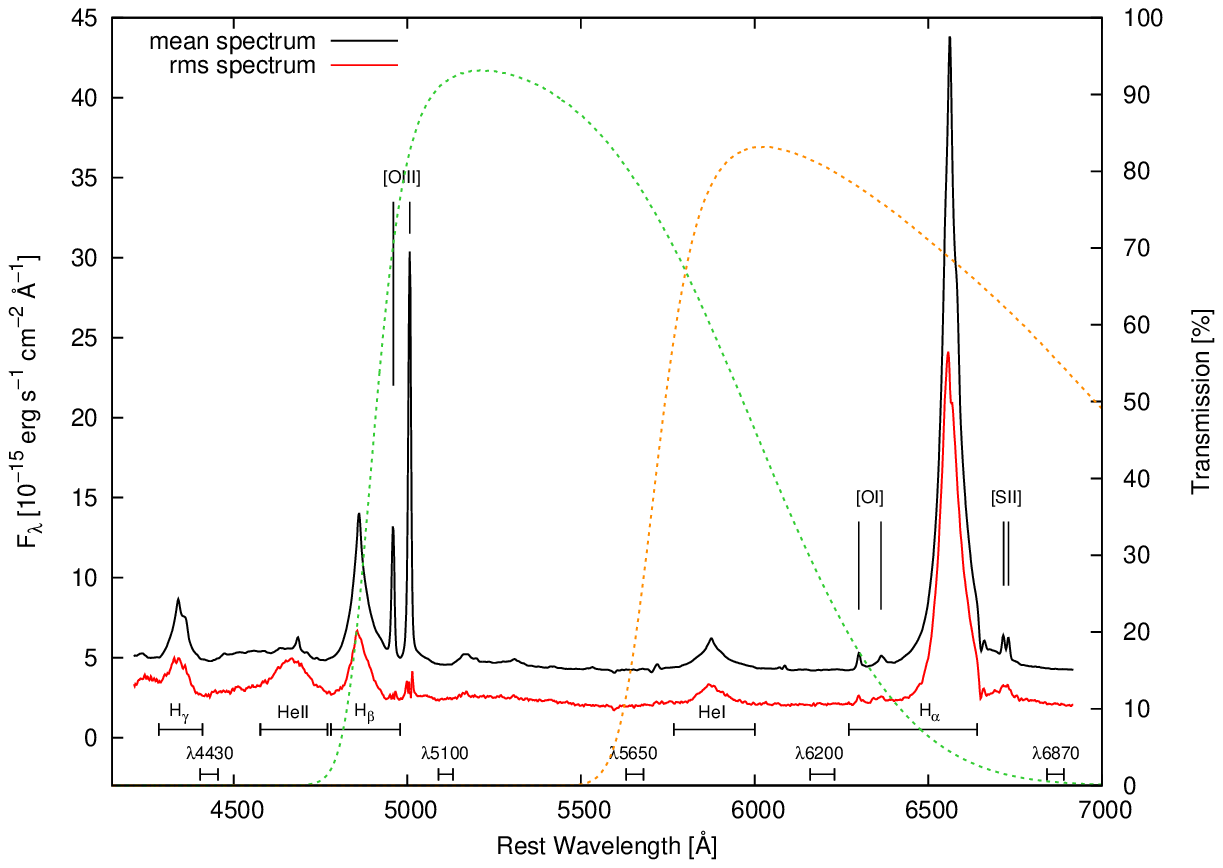}
      \caption{Integrated mean (black) and rms (red) spectra of 3C\,120.
               The rms spectrum has been scaled by a factor
               of 10 (the zero level is shifted by $-2.5$) to enhance
               weaker line structures. Overplotted are V (green) and
    R (orange) band filter curves. We also used these filters to
   generate the continuum light-curves.
              }
       \vspace*{-3mm} 
         \label{2.25_avg_rms.eps}
   \end{figure*}

\setcounter{table}{2}
\begin{table}
\centering
\tabcolsep+3.5mm
\caption{Rest frame continuum boundaries and line integration limits.}
\begin{tabular}{lccccccc}
\hline 
\noalign{\smallskip}
Cont./Line                   & Wavelength range & Pseudo-continuum \\
\noalign{\smallskip}
(1)                           & (2)                             & (3) \\
\noalign{\smallskip}
\hline 
\noalign{\smallskip}
Cont.~4430                    & 4424\,\AA{} -- 4440\,\AA{}     & \\
Cont.~5100                    & 5090\,\AA{} -- 5132\,\AA{}     & \\
Cont.~5650                    & 5630\,\AA{} -- 5680\,\AA{}     & \\
Cont.~6200                    & 6160\,\AA{} -- 6230\,\AA{}     & \\
Cont.~6870                    & 6842\,\AA{} -- 6890\,\AA{}     & \\
H$\alpha$                     & 6271\,\AA{} -- 6641\,\AA{}     & 6160\,\AA{} -- 6890\,\AA{} \\
H$\beta$                      & 4780\,\AA{} -- 4980\,\AA{}     & 4424\,\AA{} -- 5132\,\AA{} \\
H$\gamma$                     & 4284\,\AA{} -- 4410\,\AA{}     & 4424\,\AA{} -- 5132\,\AA{} \\
\ion{He}{ii}\,$\lambda 4686$  & 4577\,\AA{} -- 4770\,\AA{}     & 4424\,\AA{} -- 5132\,\AA{} \\
\ion{He}{i}\,$\lambda 5876$   & 5767\,\AA{} -- 6000\,\AA{}     & 5630\,\AA{} -- 6230\,\AA{} \\
\noalign{\smallskip}
\hline \\
\vspace{-1.1cm}
\end{tabular}
\end{table}
We present in 
Fig.~\ref{2.21_all_spectra.eps} all reduced optical spectra of 3C\,120 
that were taken during our variability campaign. All 31 spectra 
are shown in the rest frame.
They clearly show variations in the
continuum and in the \ion{He}{ii}\,$\lambda 4686$ line profile.
%
%------------------------------------------------------------------------------
%
%
%______________________________________________________________
%
Figure~\ref{2.25_avg_rms.eps} shows the mean and the rms
spectra of 3C\,120
for our variability campaign. The rms spectrum is given at the bottom.
%It is based on 3C\,120 spectra that have been scaled with respect to the
%[\ion{O}{iii}]\,$\lambda$5007 line. 
This spectrum was scaled by a factor of ten 
(the zero level is shifted by $-2.5$) to enhance weaker line structures.
The rms spectrum presents the variable part of the line profiles.
Below we discuss the spectra in the context of the
line profiles in 3C\,120 in more detail.
The continuum boundaries and line integration limits
we used for the present study are given at the bottom of the spectra in
 Figure~\ref{2.25_avg_rms.eps}. We selected the continuum boundaries
in the following way:
we inspected our mean and rms spectra for continuum regions that were
free of both strong emission and absorption lines. The final wavelength
ranges
we used for the continuum flux measurements are given in Table~3.
The continuum region at 5100 \AA\ is often used
for studies of the variable continuum flux in AGN.
In general, this region is free of strong emission lines and close
to the [\ion{O}{iii}]\,$\lambda$5007 flux calibration line.
In addition to this wavelength range at 5100 \AA\  we determined
the continuum intensities 
at four additional continuum ranges (at 4430, 5650, 6200,
and 6870 \AA{}, see Fig.\,2 and Table 3). We used these continua
to create pseudo-continua
below the variable broad emission lines as well.

We integrated the
broad emission-line intensities of the Balmer and helium lines
%\ion{He}{i}\,$\lambda 5876$
between the wavelength boundaries given in Table~3.
Figure~2 shows the selected
wavelength ranges for the Balmer and
helium lines.
First we
subtracted a linear pseudo-continuum defined by the boundaries
given in Table \,3 (Col.\,3), then we integrated the emission-line flux. 
For H$\gamma$  we extrapolated the continuum of the H$\beta$ and
\ion{He}{ii}\,$\lambda 4686$ measurements.
The results of the continuum and line intensity measurements are given in 
Table\,4.
\subsection{Continuum and emission-line light curves}
We show in Figs. 3 and 4 the continuum light-curves of 3C\,120 
 at 5100\,\AA{} and 6170\,\AA{}
for our variability campaign. 
\begin{table*}
\newcolumntype{d}{D{.}{.}{-2}}
\caption{Continuum and integrated broad-line fluxes for different epochs}
%\begin{tabular}{crrrrrrrrrr}
%\begin{tabular}{cdddddd}
\begin{tabular}{ccccccc}
\noalign{\smallskip}
\hline 
\noalign{\smallskip}
%Julian Date & 5100\,\AA  & H$\alpha$ & H$\beta$  & H$\gamma$ &   HeI     &  HeII    \\
Julian Date &   \mcr{Cont.~5100\,\AA}    & \mcc{H$\alpha$}     &   \mcc{H$\beta$}    &  \mcc{H$\gamma$}      &  \mcc{HeI}     &  \mcc{HeII}    \\
2\,450\,000+\\
(1) & (2) & (3) & (4) & (5) & (6) & (7)\\
\noalign{\smallskip}
\hline
\noalign{\smallskip}
 4726.907 & 4.045$\pm{}$ 0.099 & 26.23$\pm{}$ 0.62 & 5.21$\pm{}$ 0.08 & 1.95$\pm{}$ 0.03 & 1.71$\pm{}$ 0.04 & 1.12$\pm{}$ 0.02\\
 4729.897 & 4.300$\pm{}$ 0.038 & 25.36$\pm{}$ 0.71 & 5.49$\pm{}$ 0.05 & 1.74$\pm{}$ 0.01 & 1.67$\pm{}$ 0.05 & 1.08$\pm{}$ 0.01\\
 4739.873 & 4.187$\pm{}$ 0.278 & 24.71$\pm{}$ 2.29 & 5.01$\pm{}$ 0.32 & 2.18$\pm{}$ 0.14 & 1.49$\pm{}$ 0.14 & 0.76$\pm{}$ 0.05\\
 4740.870 & 3.891$\pm{}$ 0.030 & 22.62$\pm{}$ 0.73 & 4.62$\pm{}$ 0.06 & 1.53$\pm{}$ 0.02 & 1.46$\pm{}$ 0.05 & 1.08$\pm{}$ 0.01\\
 4747.857 & 4.345$\pm{}$ 0.033 & 26.48$\pm{}$ 0.44 & 5.47$\pm{}$ 0.07 & 1.81$\pm{}$ 0.02 & 1.77$\pm{}$ 0.03 & 1.17$\pm{}$ 0.01\\
 4748.847 & 4.117$\pm{}$ 0.035 & 25.19$\pm{}$ 0.37 & 5.12$\pm{}$ 0.05 & 1.70$\pm{}$ 0.02 & 1.64$\pm{}$ 0.02 & 1.10$\pm{}$ 0.01\\
 4749.850 & 4.229$\pm{}$ 0.043 & 25.59$\pm{}$ 0.36 & 5.30$\pm{}$ 0.03 & 1.74$\pm{}$ 0.01 & 1.68$\pm{}$ 0.02 & 1.13$\pm{}$ 0.01\\
 4752.834 & 4.381$\pm{}$ 0.057 & 25.46$\pm{}$ 0.93 & 5.52$\pm{}$ 0.06 & 1.77$\pm{}$ 0.02 & 1.73$\pm{}$ 0.06 & 1.07$\pm{}$ 0.01\\
 4759.967 & 4.318$\pm{}$ 0.040 & 24.86$\pm{}$ 0.52 & 5.41$\pm{}$ 0.07 & 1.75$\pm{}$ 0.02 & 1.66$\pm{}$ 0.04 & 1.04$\pm{}$ 0.01\\
 4762.815 & 3.790$\pm{}$ 0.265 & 23.35$\pm{}$ 2.20 & 4.81$\pm{}$ 0.42 & 1.82$\pm{}$ 0.16 & 1.49$\pm{}$ 0.14 & 0.79$\pm{}$ 0.07\\
 4771.774 & 4.219$\pm{}$ 0.080 & 23.43$\pm{}$ 0.47 & 4.78$\pm{}$ 0.14 & 1.79$\pm{}$ 0.05 & 1.40$\pm{}$ 0.03 & 0.77$\pm{}$ 0.02\\
 4775.771 & 4.445$\pm{}$ 0.026 & 25.25$\pm{}$ 0.49 & 4.92$\pm{}$ 0.03 & 1.62$\pm{}$ 0.01 & 1.67$\pm{}$ 0.03 & 0.90$\pm{}$ 0.01\\
 4776.787 & 4.272$\pm{}$ 0.044 & 24.95$\pm{}$ 0.46 & 4.81$\pm{}$ 0.06 & 1.55$\pm{}$ 0.02 & 1.66$\pm{}$ 0.03 & 0.91$\pm{}$ 0.01\\
 4779.773 & 4.660$\pm{}$ 0.045 & 26.21$\pm{}$ 0.57 & 5.18$\pm{}$ 0.07 & 1.62$\pm{}$ 0.02 & 1.75$\pm{}$ 0.04 & 0.93$\pm{}$ 0.01\\
 4783.905 & 4.297$\pm{}$ 0.037 & 24.17$\pm{}$ 0.70 & 4.72$\pm{}$ 0.03 & 1.44$\pm{}$ 0.01 & 1.64$\pm{}$ 0.05 & 0.98$\pm{}$ 0.01\\
 4787.745 & 4.918$\pm{}$ 0.045 & 25.88$\pm{}$ 0.36 & 5.11$\pm{}$ 0.03 & 1.59$\pm{}$ 0.01 & 1.73$\pm{}$ 0.02 & 0.93$\pm{}$ 0.01\\
 4789.737 & 5.065$\pm{}$ 0.274 & 25.13$\pm{}$ 1.74 & 5.08$\pm{}$ 0.30 & 1.51$\pm{}$ 0.09 & 1.76$\pm{}$ 0.12 & 1.00$\pm{}$ 0.06\\
 4790.736 & 5.052$\pm{}$ 0.033 & 25.76$\pm{}$ 0.66 & 4.91$\pm{}$ 0.03 & 1.56$\pm{}$ 0.01 & 1.68$\pm{}$ 0.04 & 0.90$\pm{}$ 0.01\\
 4793.723 & 5.073$\pm{}$ 0.034 & 24.46$\pm{}$ 0.34 & 4.92$\pm{}$ 0.01 & 1.60$\pm{}$ 0.01 & 1.60$\pm{}$ 0.02 & 1.15$\pm{}$ 0.01\\
 4807.681 & 5.819$\pm{}$ 0.311 & 26.28$\pm{}$ 1.81 & 5.39$\pm{}$ 0.30 & 1.81$\pm{}$ 0.10 & 1.81$\pm{}$ 0.12 & 1.65$\pm{}$ 0.09\\
 4808.828 & 5.615$\pm{}$ 0.043 & 24.73$\pm{}$ 0.56 & 5.58$\pm{}$ 0.05 & 2.49$\pm{}$ 0.02 & 1.70$\pm{}$ 0.04 & 2.16$\pm{}$ 0.02\\
 4822.652 & 4.824$\pm{}$ 0.039 & 22.69$\pm{}$ 0.74 & 5.16$\pm{}$ 0.16 & 2.29$\pm{}$ 0.05 & 1.71$\pm{}$ 0.06 & 1.79$\pm{}$ 0.04\\
 4830.622 & 4.859$\pm{}$ 0.060 & 25.62$\pm{}$ 0.66 & 5.83$\pm{}$ 0.10 & 2.59$\pm{}$ 0.04 & 1.82$\pm{}$ 0.05 & 1.72$\pm{}$ 0.03\\
 4841.600 & 4.910$\pm{}$ 0.040 & 25.59$\pm{}$ 0.47 & 5.99$\pm{}$ 0.06 & 2.67$\pm{}$ 0.03 & 1.84$\pm{}$ 0.03 & 1.65$\pm{}$ 0.02\\
 4845.725 & 4.532$\pm{}$ 0.041 & 24.02$\pm{}$ 0.63 & 5.28$\pm{}$ 0.05 & 2.19$\pm{}$ 0.02 & 1.67$\pm{}$ 0.04 & 1.39$\pm{}$ 0.01\\
 4866.691 & 4.261$\pm{}$ 0.257 & 21.74$\pm{}$ 1.54 & 4.88$\pm{}$ 0.29 & 2.06$\pm{}$ 0.12 & 1.54$\pm{}$ 0.11 & 1.43$\pm{}$ 0.09\\
 4876.637 & 4.829$\pm{}$ 0.085 & 24.77$\pm{}$ 0.90 & 5.51$\pm{}$ 0.12 & 2.37$\pm{}$ 0.05 & 1.65$\pm{}$ 0.05 & 1.75$\pm{}$ 0.04\\
 4881.635 & 5.061$\pm{}$ 0.025 & 25.31$\pm{}$ 0.55 & 5.97$\pm{}$ 0.09 & 2.52$\pm{}$ 0.04 & 1.74$\pm{}$ 0.04 & 1.88$\pm{}$ 0.03\\
 4887.617 & 4.916$\pm{}$ 0.036 & 25.74$\pm{}$ 0.53 & 5.59$\pm{}$ 0.10 & 1.79$\pm{}$ 0.03 & 1.78$\pm{}$ 0.04 & 1.35$\pm{}$ 0.02\\
 4893.604 & 4.722$\pm{}$ 0.043 & 25.46$\pm{}$ 0.55 & 5.79$\pm{}$ 0.05 & 2.09$\pm{}$ 0.02 & 1.69$\pm{}$ 0.04 & 1.45$\pm{}$ 0.01\\
 4906.576 & 3.807$\pm{}$ 0.036 & 23.49$\pm{}$ 0.39 & 5.23$\pm{}$ 0.07 & 1.82$\pm{}$ 0.02 & 1.67$\pm{}$ 0.03 & 1.00$\pm{}$ 0.01\\
\noalign{\smallskip}
\hline 
\noalign{\smallskip}
\end{tabular}

Continuum flux (2) in units of 10$^{-15}$\,erg\,s$^{-1}$\,cm$^{-2}$\,\AA$^{-1}$.\\
Line fluxes (3) - (7) in units 10$^{-13}$\,erg\,s$^{-1}$\,cm$^{-2}$.
\end{table*}
%new
\onecolumn
\tabcolsep+4mm
\newcolumntype{d}{D{.}{.}{-1}}
\begin{longtable}{ldldc}
\caption{Continuum fluxes at 5100\,\AA{} (V-band) and 6170\,\AA{} (R-band) taken
with the HET (H), Wise (W), and Cerro Armazones (C) telescopes at different
epochs.}\\
  \hline 
 %\noalign{\smallskip}
 Julian Date & \mcr{Cont. Flux} & Julian Date & \mcr{Cont. Flux}&Telescope\\
 2\,450\,000+& \mcr{5100\,\AA}  & 2\,450\,000+& \mcr{6200\,\AA} &         \\
 %(1) & (2) & (3) & (4) \\ %& (5) & (6) & (7) & (8) \\ 
 %\noalign{\smallskip}
 \hline 
 %\noalign{\smallskip}    
  \endfirsthead
\caption{continued.}\\
  \hline 
 %\noalign{\smallskip}
 Julian Date & \mcr{Cont. Flux} & Julian Date & \mcr{Cont. Flux}&Telescope\\
 2\,450\,000+& \mcr{5100\,\AA}  & 2\,450\,000+& \mcr{6200\,\AA} &         \\
 %(1) & (2) & (3) & (4) \\ %& (5) & (6) & (7) & (8) \\ 
 %\noalign{\smallskip}
 \hline 
 %\noalign{\smallskip} 
\endhead
\hline
\multicolumn{3}{l}{Continuum flux in units of  10$^{-15}$\,erg\,s$^{-1}$\,cm$^{-2}$\,\AA$^{-1}$.}\\
\endfoot
 4726.907 & 4.045\pm 0.099 &    4726.907 & 4.158\pm 0.056 & H  \\
 4729.897 & 4.300\pm 0.038 &    4729.897 & 3.967\pm 0.035 & H  \\
 4739.873 & 4.187\pm 0.278 &    4739.873 & 4.076\pm 0.287 & H  \\
 4740.870 & 3.891\pm 0.030 &    4740.870 & 3.765\pm 0.044 & H  \\
 4745.580 & 4.129\pm 0.111 &    4745.576 & 3.848\pm 0.099 & W  \\
 4745.588 & 4.163\pm 0.104 &    4745.584 & 3.805\pm 0.095 & W  \\
 4747.857 & 4.345\pm 0.033 &    4747.857 & 4.161\pm 0.029 & H  \\
 4748.847 & 4.117\pm 0.035 &    4748.847 & 3.958\pm 0.030 & H  \\
 4749.850 & 4.229\pm 0.043 &    4749.850 & 4.004\pm 0.026 & H  \\
 4752.834 & 4.381\pm 0.057 &    4752.834 & 3.951\pm 0.029 & H  \\
 4759.967 & 4.318\pm 0.040 &    4759.967 & 3.907\pm 0.045 & H  \\
 4762.815 & 3.790\pm 0.265 &    4762.815 & 3.716\pm 0.266 & H  \\
 4771.774 & 4.219\pm 0.080 &    4771.774 & 4.084\pm 0.076 & H  \\
 4775.771 & 4.445\pm 0.026 &    4775.771 & 4.176\pm 0.031 & H  \\
 4776.787 & 4.272\pm 0.044 &    4776.787 & 4.106\pm 0.028 & H  \\
 4779.773 & 4.660\pm 0.045 &    4779.773 & 4.337\pm 0.033 & H  \\
 4783.905 & 4.297\pm 0.037 &    4783.905 & 4.083\pm 0.052 & H  \\
 4787.745 & 4.918\pm 0.045 &    4787.745 & 4.464\pm 0.030 & H  \\
          &                &    4788.625 & 4.406\pm 0.088 & C  \\
 4789.737 & 5.065\pm 0.274 &    4789.737 & 4.514\pm 0.250 & H  \\
          &                &    4790.625 & 4.470\pm 0.094 & C  \\
 4790.736 & 5.052\pm 0.033 &    4790.736 & 4.589\pm 0.051 & H  \\
          &                &    4791.625 & 4.488\pm 0.096 & C  \\
          &                &    4792.625 & 4.455\pm 0.093 & C  \\
          &                &    4793.625 & 4.456\pm 0.093 & C  \\
 4793.723 & 5.073\pm 0.034 &    4793.723 & 4.431\pm 0.031 & H  \\
          &                &    4794.625 & 4.539\pm 0.101 & C  \\
          &                &    4796.625 & 4.504\pm 0.098 & C  \\
 4797.308 & 5.220\pm 0.118 &    4797.304 & 4.456\pm 0.111 & W  \\
          &                &    4797.625 & 4.442\pm 0.092 & C  \\
          &                &    4798.625 & 4.505\pm 0.098 & C  \\
          &                &    4799.625 & 4.532\pm 0.101 & C  \\
          &                &    4800.625 & 4.510\pm 0.099 & C  \\
          &                &    4802.625 & 4.505\pm 0.098 & C  \\
 4803.301 & 5.597\pm 0.129 &    4803.297 & 4.798\pm 0.096 & W  \\
 4803.309 & 5.562\pm 0.122 &    4803.305 & 4.767\pm 0.101 & W  \\
          &                &    4803.625 & 4.528\pm 0.100 & C  \\
 4804.308 & 5.486\pm 0.151 &    4804.304 & 4.655\pm 0.113 & W  \\
 4804.315 & 5.513\pm 0.148 &    4804.312 & 4.643\pm 0.116 & W  \\
 4804.326 & 5.507\pm 0.152 &    4804.322 & 4.692\pm 0.121 & W  \\
          &                &    4804.331 & 4.705\pm 0.114 & W  \\ 
          &                &    4804.625 & 4.557\pm 0.103 & C  \\
 4805.299 & 5.633\pm 0.121 &    4805.296 & 4.792\pm 0.104 & W  \\
 4805.307 & 5.604\pm 0.136 &    4805.303 & 4.817\pm 0.105 & W  \\
 4805.335 & 5.562\pm 0.135 &    4805.331 & 4.730\pm 0.103 & W  \\
 4807.681 & 5.819\pm 0.311 &    4807.681 & 5.068\pm 0.289 & H  \\
 4808.346 & 5.500\pm 0.124 &    4808.342 & 4.855\pm 0.103 & W  \\
 4808.354 & 5.689\pm 0.131 &    4808.350 & 4.887\pm 0.119 & W  \\
 4808.361 & 5.486\pm 0.130 &    4808.357 & 4.849\pm 0.100 & W  \\ 
 4808.828 & 5.615\pm 0.043 &    4808.828 & 4.866\pm 0.040 & H  \\
 4811.331 & 5.725\pm 0.683 &    4811.327 & 4.983\pm 0.870 & W  \\
 4817.286 & 5.640\pm 0.159 &    4817.282 & 5.074\pm 0.127 & W  \\
 4817.512 & 5.576\pm 0.178 &    4817.508 & 5.022\pm 0.140 & W  \\
 4818.293 & 5.437\pm 0.154 &    4818.289 & 4.874\pm 0.163 & W  \\
 4819.301 & 5.362\pm 0.121 &    4819.297 & 4.680\pm 0.099 & W  \\ 
 4822.652 & 4.824\pm 0.039 &    4822.652 & 4.466\pm 0.034 & H  \\
          &                &    4825.353 & 4.944\pm 0.131 & W  \\
          &                &    4826.432 & 4.849\pm 0.115 & W  \\
 4830.622 & 4.859\pm 0.060 &    4830.622 & 4.615\pm 0.047 & H  \\
 4837.209 & 4.600\pm 0.147 &    4837.205 & 4.245\pm 0.128 & W  \\
 4837.217 & 4.687\pm 0.165 &    4837.213 & 4.176\pm 0.120 & W  \\
          &                &    4838.203 & 4.408\pm 0.216 & W  \\
 4839.335 & 4.849\pm 0.349 &    4839.332 & 4.396\pm 0.252 & W  \\ 
 4841.600 & 4.910\pm 0.040 &    4841.600 & 4.524\pm 0.024 & H  \\
 4845.725 & 4.532\pm 0.041 &    4845.725 & 4.284\pm 0.032 & H  \\
 4852.176 & 5.003\pm 0.118 &    4852.172 & 4.332\pm 0.105 & W  \\
 4853.174 & 4.786\pm 0.119 &    4853.170 & 4.181\pm 0.096 & W  \\
 4856.205 & 5.281\pm 0.135 &    4856.201 & 4.680\pm 0.108 & W  \\
 4859.182 & 4.856\pm 0.118 &    4859.178 & 4.227\pm 0.109 & W  \\
 4866.195 & 4.805\pm 0.269 &    4866.192 & 4.268\pm 0.172 & W  \\
 4866.691 & 4.261\pm 0.257 &    4866.691 & 3.951\pm 0.225 & H  \\
 4867.184 & 4.749\pm 0.186 &    4867.180 & 4.216\pm 0.226 & W  \\ 
 4876.637 & 4.829\pm 0.085 &    4876.637 & 4.435\pm 0.091 & H  \\
 4881.635 & 5.061\pm 0.025 &    4881.635 & 4.422\pm 0.038 & H  \\
 4887.617 & 4.916\pm 0.036 &    4887.617 & 4.336\pm 0.029 & H  \\
 4893.604 & 4.722\pm 0.043 &    4893.604 & 4.337\pm 0.032 & H  \\
 4894.195 & 4.426\pm 0.122 &    4894.191 & 4.035\pm 0.095 & W  \\
 4895.196 & 4.366\pm 0.114 &    4895.192 & 3.952\pm 0.110 & W  \\
          &                &    4903.304 & 3.974\pm 0.184 & W  \\
 4905.209 & 3.939\pm 0.106 &    4905.205 & 3.667\pm 0.087 & W  \\ 
 4906.576 & 3.807\pm 0.036 &    4906.576 & 3.687\pm 0.047 & H  \\
 4908.242 & 3.884\pm 0.094 &    4908.238 & 3.704\pm 0.083 & W  \\
 4920.219 & 3.681\pm 0.087 &    4920.215 & 3.482\pm 0.076 & W  \\
 5064.531 & 2.944\pm 0.081 &    5064.527 & 2.737\pm 0.074 & W  \\
 5064.538 & 2.912\pm 0.082 &    5064.535 & 2.768\pm 0.073 & W  \\
 5098.574 & 2.386\pm 0.098 &    5098.571 & 2.395\pm 0.076 & W  \\
 5117.536 & 2.953\pm 0.088 &    5117.533 & 2.599\pm 0.075 & W  \\ 
\end{longtable}
\twocolumn

%----------------------------------------------------------------------------- 
%                                                
   \begin{figure*}
    \includegraphics[width=15cm,angle=0]{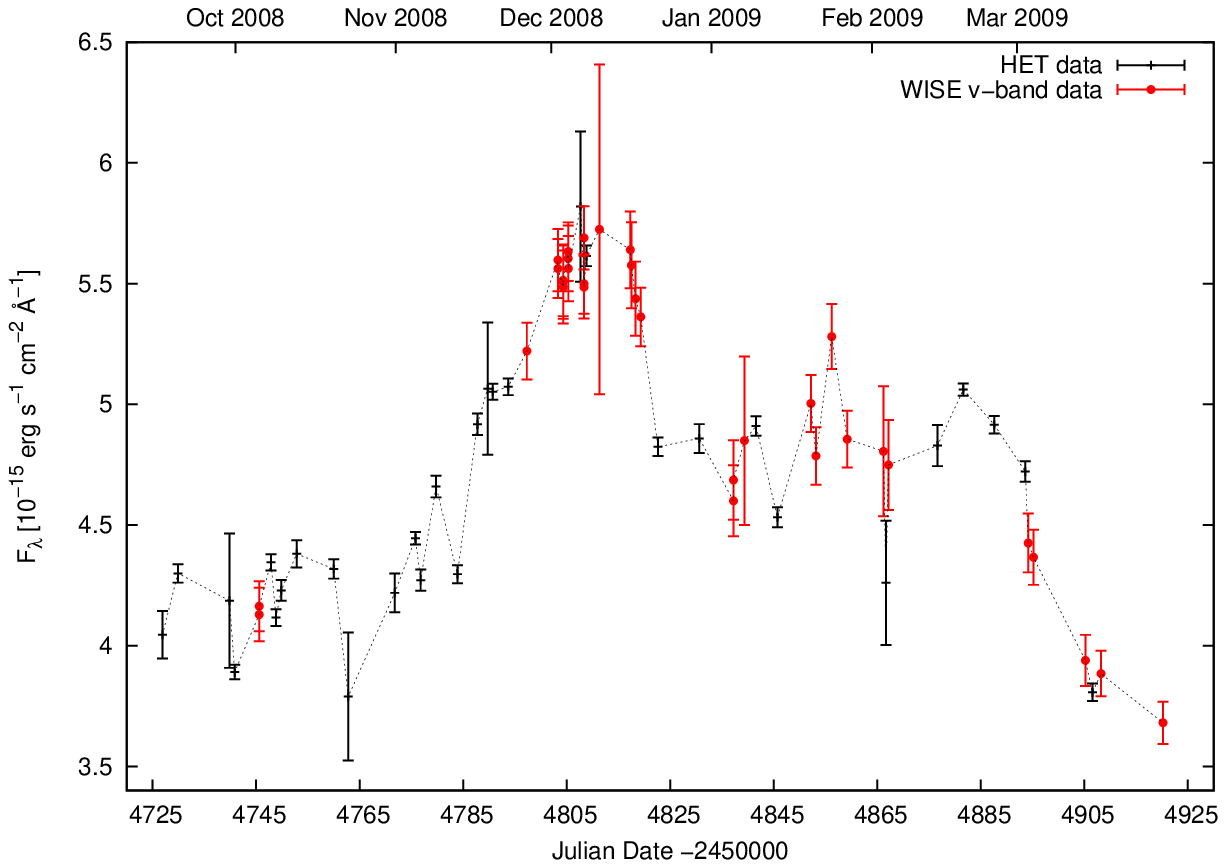}
      \caption{Continuum light-curve at 5100\,\AA\ (including V-band data)
 for 3C\,120}
% for our variability campaign HET 5100, Wise v-band data.}
         \label{3.2_lightcurve_5100.eps}
%   \end{figure*}
%
%______________________________________________________________
%
%----------------------------------------------------------------------------- 
%                                                
%   \begin{figure*}
%    \includegraphics[bb=40 90 380 700,width=9.12cm,angle=-90]{2.25_avg_rms.eps}
%    \includegraphics[width=14cm,angle=0]{3.3_lightcurve_6170.eps}
%    \includegraphics[width=14cm,angle=0]{Fig4_lightcurve6170_neu.eps}
    \includegraphics[width=16cm,angle=0]{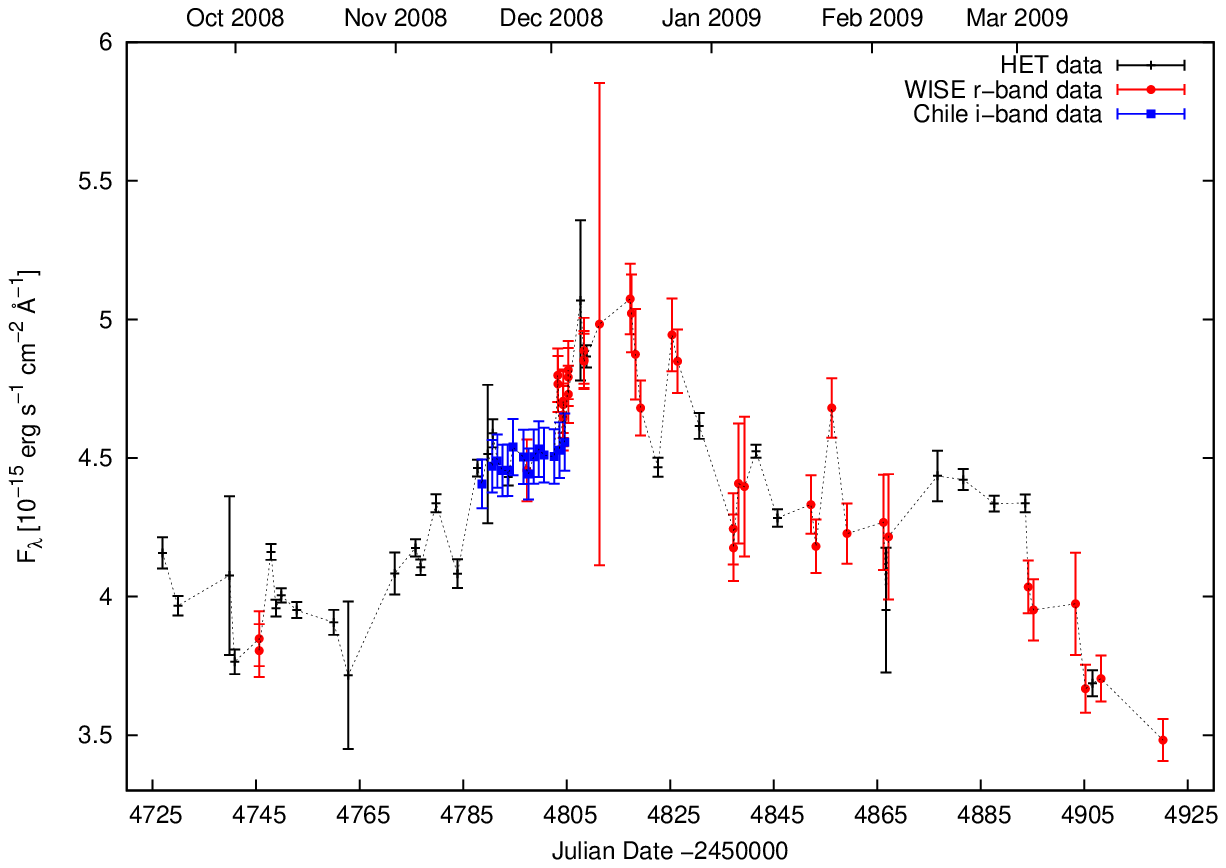}
     \caption{Continuum light-curve at 6200\,\AA\ (including R-band, and
intercalibrated I-band data) for 3C\,120}
%      \caption{Continuum light curve of 3C\,120 for our variability campaign.
%               HET 6170, Wise r-band, Chile i-band data.}
         \label{3.3_lightcurve_6170.eps}
   \end{figure*}
%
%______________________________________________________________
%\clearpage
%

%----------------------------------------------------------------------------- 
%                                                
   \begin{figure*}[t]
\centering
    \includegraphics[width=15cm,angle=0]{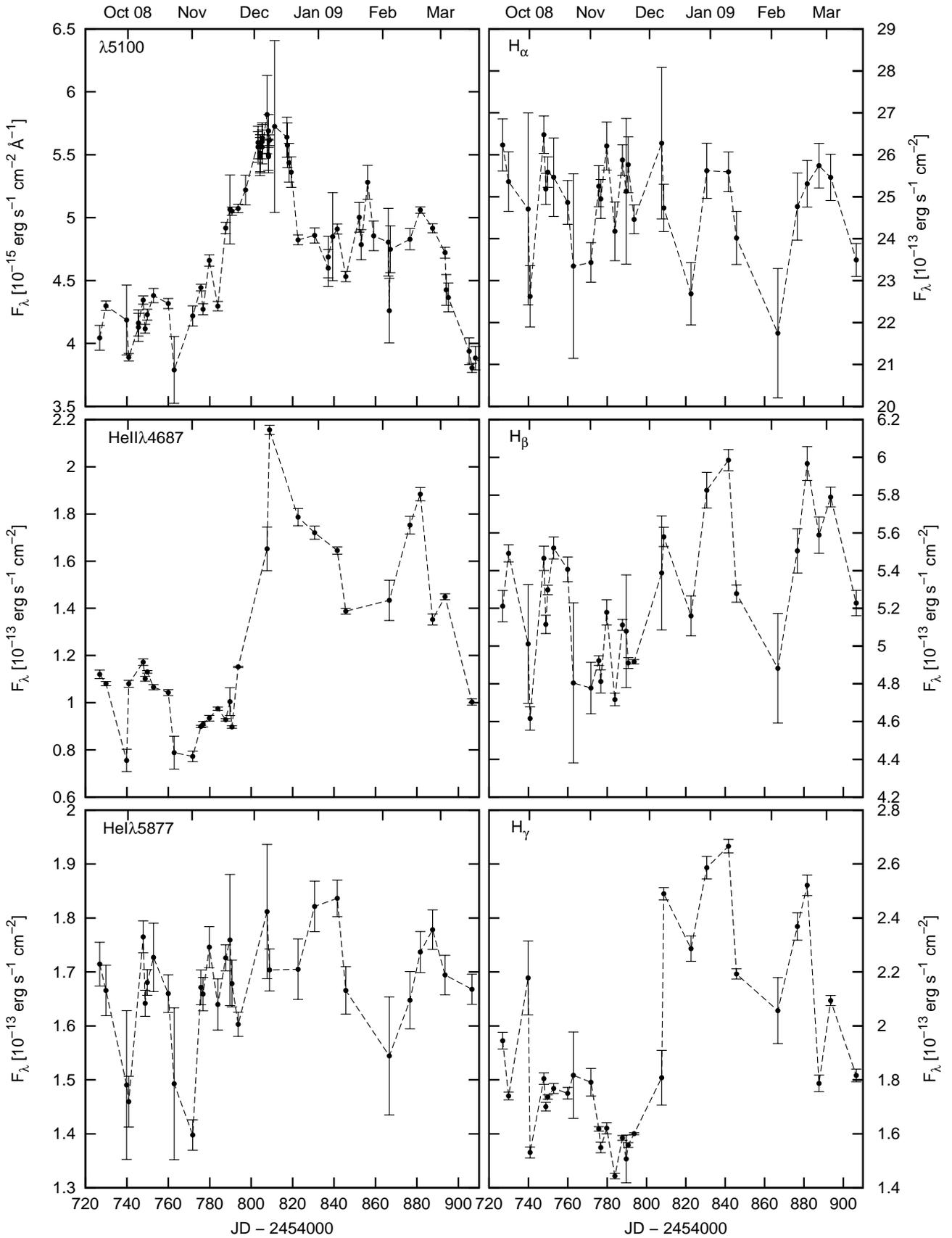}
\vspace{1.5cm}
      \caption{Light curves of the continuum flux at 5100\,\AA\
 (in units of 10$^{-15}$ erg cm$^{-2}$ s$^{-1}$\,\AA$^{-1}$) and of
the integrated emission-line fluxes of H$\alpha$, H$\beta$,
H$\gamma$, \ion{He}{ii}\,$\lambda 4686$ , and \ion{He}{i}\,$\lambda 5876$
   (in units of  10$^{-13}$ erg cm$^{-2}$ s$^{-1}$).}
         \label{3.5_linelightcurves.eps}
   \end{figure*}
%
%______________________________________________________________
%
%
\clearpage
First we created light curves based
on the HET spectra alone. Then we generated light curves based on the
V- and R-band photometric data taken at Wise observatory. We
intercalibrated the V-band photometry into the continuum light curve
at 5100\,\AA{} and the R-band photometry into the continuum light curve
at 6170\,\AA.
We applied a multiplicative scale factor and an additional flux adjustment
component to set the light curves on the same scale and to correct
for differences in the host galaxy contribution.
 Finally, we also fitted the I'-band photometry observed at
Cerro Armazones into the continuum light curve at 6170\,\AA.
Overall, the light curves from different telescopes agree well.
 The individual continuum fluxes for the different epochs
at 5100\,\AA{} and 6170\,\AA{} are given in Table 5.
 The light curve in the V band has a higher
variation amplitude than the light curve in the R band, as expected for
AGN where the contribution of the nonthermal continuum flux is stronger
in the blue.

%(Shai-Info zu Lichtkurven:
%he directory has in it the light curves of 3C120 observed during 2008 to 2009.
%The files star.dat are light curve extracted from spectroscopy of HET
% (sent to me by Matthias).\\ 
%To these light curves I fitted the photometry from Wise:
%The R band photometry was intercalibrated into the f6170.dat file and it 
%is in the lc-phot-R file.
%The V band photometry was intercalibrated into the f5100.dat file and it 
%is in the lc-phot-V file.\\
%To the f6170.dat file I fitted also the I band photometry observed
%by Martin Hass. The intercalibrated data are in the lc-phot-I file.\\
%Final light curves are:
%lc-5100 - combined the data from f5100.dat and lc-phot-V.
%lc-6170 - combined the data from f6170.dat, lc-phot-R, and lc-phot-I.\\
%Final light curves are presented in the star.eps files. 
%The files with star-a.eps show the light curve during the monitoring campaign,
%the other files show the whole light curve including three photometric points
% taken  year later.\\
%
%\onecolumn

%\clearpage
In Fig. 5 we present the light curves of the integrated
emission-line fluxes of the Balmer lines  H$\alpha$, H$\beta$, and H$\gamma$,
and of the of the helium lines
\ion{He}{i}\,$\lambda 5876$,  \ion{He}{ii}\,$\lambda 4686$.
The continuum light curve at 5100\,\AA{} is shown for comparison.
 The line and continuum flux values are given in Table 4.
The mean continuum flux
% of 2.823$\pm$0.229 
$F_{\lambda}(5100$\,\AA$)$ is  $4.57\pm0.23 \times 
10^{-15}$\,erg\,s$^{-1}$\,cm$^{-2}$\,\AA$^{-1}$
%.
% This corresponds to a mean luminosity of
%$L(5180$\,\AA$) = 1.159\pm0.094 \times 10^{40}$\,erg\,s$^{-1}$\,\AA$^{-1}$ or
%for $H_{0}$=70km/s/Mpc and\\
%$\lambda$\,$L_{\lambda}(5180$\,\AA$) = 5.91\pm0.56 \times
%10^{43}$\,erg\,s$^{-1}$  (Daten, Fehler checken und mit Benz vergleichen).  
and the mean H$\beta$ flux 
F(H$\beta)$ is 5.24$\pm0.39 \times 10^{-13}$\,erg\,s$^{-1}$\,cm$^{-2}$.
%and the mean H$\beta$ luminosity to
%L(H$\beta$) = 1.59$\pm0.16 \times 10^{42}$\,erg\,s$^{-1}$\,\AA$^{-1}$.
%% $L(H\beta$) = 1.59$\pm0.15984 \times
%In Fig.~3 and 4  we present the light curves of the continuum fluxes at
%5100 and 6170\,\AA\
% (in units of 10$^{-15}$ erg cm$^{-2}$ s$^{-1}$\,\AA$^{-1}$) as well as
%the light curves of the integrated
%  emission-line fluxes of  the Balmer lines H$\alpha$, H$\beta$, and H$\gamma$
%and of the Helium lines \ion{He}{ii}\,$\lambda 4686$ ,
%and \ion{He}{i}\,$\lambda 5876$
% (in units of  10$^{-15}$ erg cm$^{-2}$ s$^{-1}$).
%
%\clearpage

Some statistics of the continuum and emission line intensity variations
is given in Table\,6.
We indicate the lowest and highest fluxes F$_{min}$ and F$_{max}$,
peak-to-peak amplitudes R$_{max}$ = F$_{max}$/F$_{min}$, the mean flux
during the period of observations $<$F$>$, the standard deviation $\sigma_F$,
 and the fractional variation 
\[ F_{var} = \frac{\sqrt{{\sigma_F}^2 - \Delta^2}}{<F>} \] 
as defined by Rodr\'\i{}guez-Pascual et al.\cite{rodriguez97}.
The quantity $\Delta^2$ is the mean square value of the uncertainties 
$\Delta_{i}$ associated with the fluxes $F_{i}$.
\begin{table}
\centering
\tabcolsep+1mm
\caption{Variability statistics for 3C\,120 in units of
 10$^{-15}$\,erg\,s$^{-1}$\,cm$^{-2}$\,\AA$^{-1}$ for the continuum
(columns 2, 3, 5) and in units of 
 10$^{-15}$\,erg\,s$^{-1}$\,cm$^{-2}$ for the emission lines
(columns 2, 3, 5).}
\begin{tabular}{lccccccc}
\hline 
\noalign{\smallskip}
Cont./Line & F$_{min}$ & F$_{max}$ & R$_{max}$ & $<$F$>$ & $\sigma_F$ & F$_{var
}$ \\
\noalign{\smallskip}
(1) & (2) & (3) & (4) & (5) & (6) & (7) \\ %(8) & (9) \\ 
\noalign{\smallskip}
\hline 
\noalign{\smallskip}
%Cont.~5100                    &    3.790 &    5.819 & 1.536   &    4.573 &   0.491    & 0.104   \\
%H$\alpha$                     & 2174.481 & 2647.703 & 1.218   & 2485.562 & 117.072    & 0.028   \\
%H$\beta$                      &  461.633 &  598.527 & 1.297   &  524.333 &  36.940    & 0.065   \\
%H$\gamma$                     &  144.357 &  266.614 & 1.847   &  190.054 &  34.937    & 0.181   \\
%\ion{He}{i}\,$\lambda 5876$   &  139.765 &  183.634 & 1.314   &  167.066 &  10.423    & 0.050   \\
%\ion{He}{ii}\,$\lambda 4686$  &   75.557 &  215.628 & 2.854   &  122.685 &  36.712    & 0.298   \\
Cont.~5100                    &    3.79 &    5.82 & 1.54   &    4.57 &   0.491   & 0.104   \\
Cont.~6200                    &    3.69 &    5.07 & 1.37    &    4.24 &   0.32    & 0.071 \\ 
H$\alpha$                     & 2174.5  & 2647.7  & 1.22   & 2485.6  & 117.07    & 0.028   \\
%H$\alpha$                     & 2174.48 & 2647.70 & 1.218   & 2485.56 & 117.07    & 0.028   \\
H$\beta$                      &  461.6 &  598.5 & 1.30   &  524.3 &  36.94    & 0.065   \\
H$\gamma$                     &  144.4 &  266.6 & 1.85   &  190.1 &  34.94    & 0.181   \\
\ion{He}{i}\,$\lambda 5876$   &  139.8 &  183.6 & 1.31   &  167.1 &  10.42    & 0.050   \\
\ion{He}{ii}\,$\lambda 4686$  &   75.6 &  215.6 & 2.85   &  122.7 &  36.71    & 0.298   \\
\noalign{\smallskip}          
\hline 
\end{tabular}
\end{table}

%
%The intrinsic variations are of the order of 7 to 15\%.  
%Hier ist noch nicht die Hostgalaxie abgezogen (Wert aus Bentz nehmen?).

\subsection{Mean and rms line profiles}
%------------------------------------------------------------------------------
Based on the observed spectra, we calculated
normalized mean and rms
line profiles of the Balmer lines H$\alpha$,
 H$\beta$,  H$\gamma$, and of the helium lines
\ion{He}{i}\,$\lambda 5876$,  and \ion{He}{ii}\,$\lambda 4686$
after subtracting the continuum flux.  
\begin{figure}[t]
 \includegraphics[width=9cm]{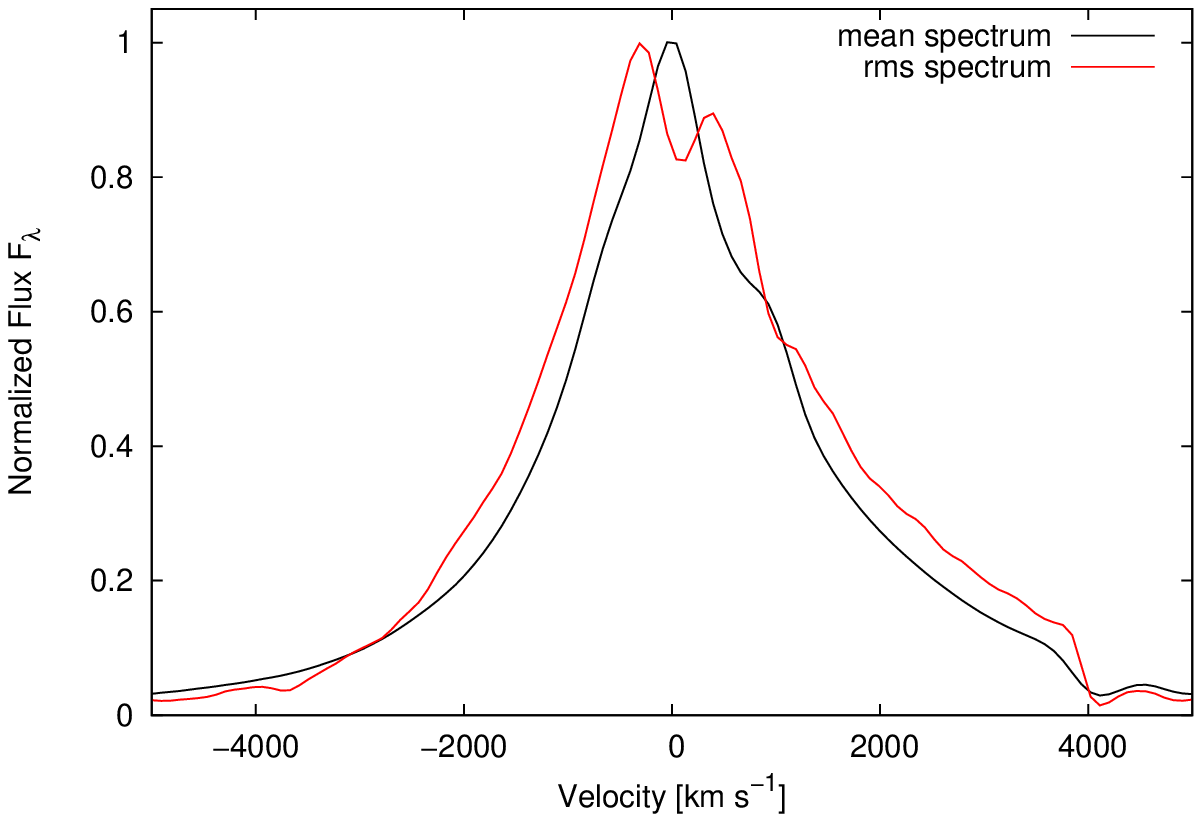}
  \caption{Normalized mean (black) and rms (red) line
 profiles of H$\alpha$
   in velocity space.}
  \label{ha0405meanrms.ps}
%\end{figure}
%
%\begin{figure}
%  \includegraphics[width=9cm]{Fig9_avgrms_hb.eps}
  \includegraphics[width=9cm]{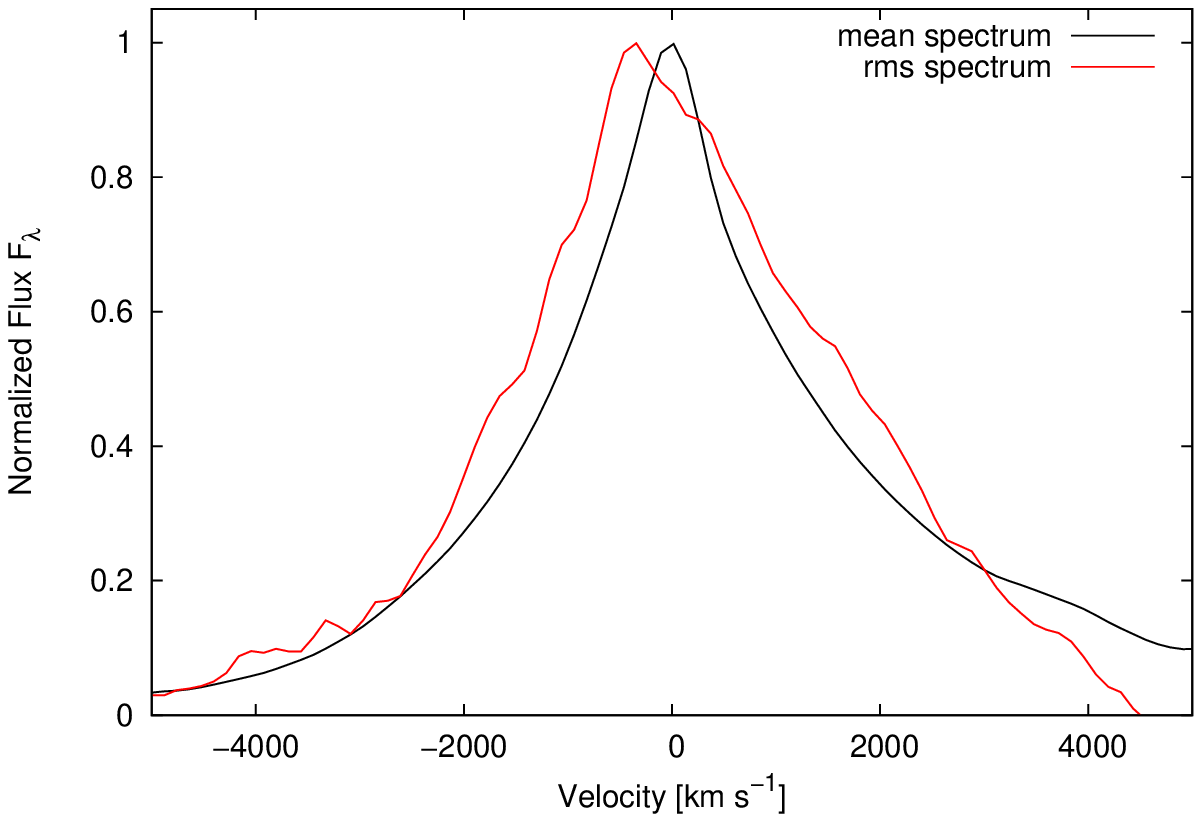}
  \caption{Normalized mean (black) and rms (red)
 line profiles of H$\beta$ in velocity space.}
  \label{hb0405meanrms.ps}
%\end{figure}
%
%\begin{figure}[b]
%  \includegraphics[width=9cm]{Fig10_avgrms_hg.eps}
  \includegraphics[width=9cm]{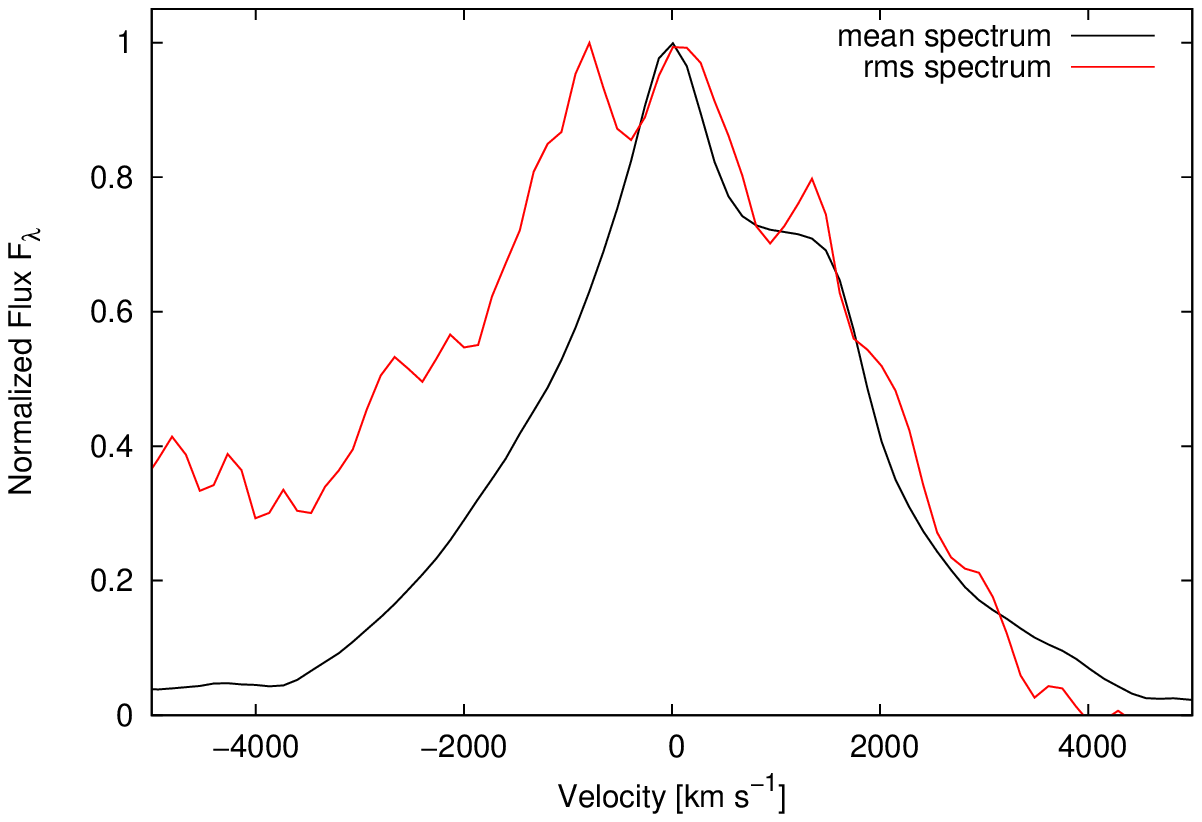}
  \caption{Normalized mean (black) and rms (red)
 line profiles of H$\gamma$ in velocity space. The strong blue wing in the
rms profile is probably due to a poorly accounted continuum (as described
 in the text).}
  \label{hb0405meanrms.ps}
\end{figure}
\begin{figure}
  \includegraphics[width=9cm]{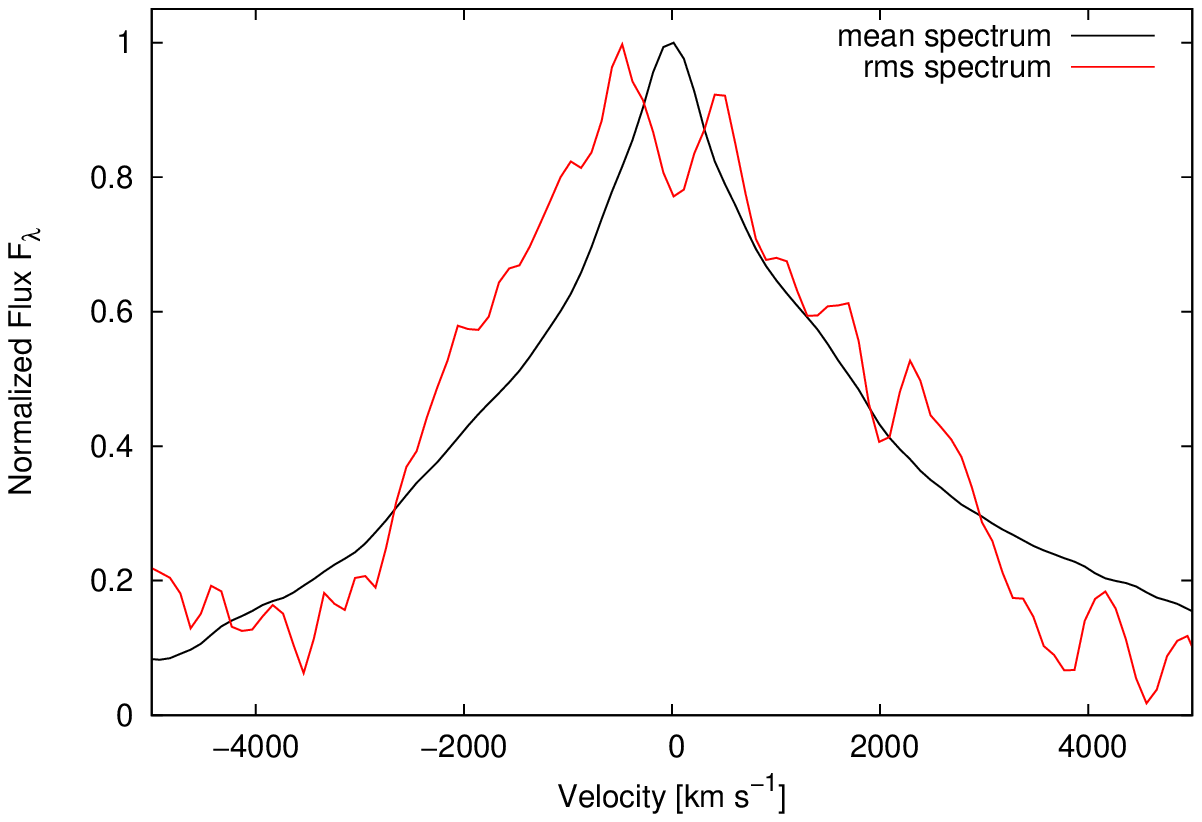}
  \caption{Normalized mean (black) and rms (red)
 line profiles of \ion{He}{i}\,$\lambda 5876$ in velocity space.}
  \label{hb0405meanrms.ps}
\end{figure}
%
%\clearpage
%
\begin{figure}
  \includegraphics[width=9cm]{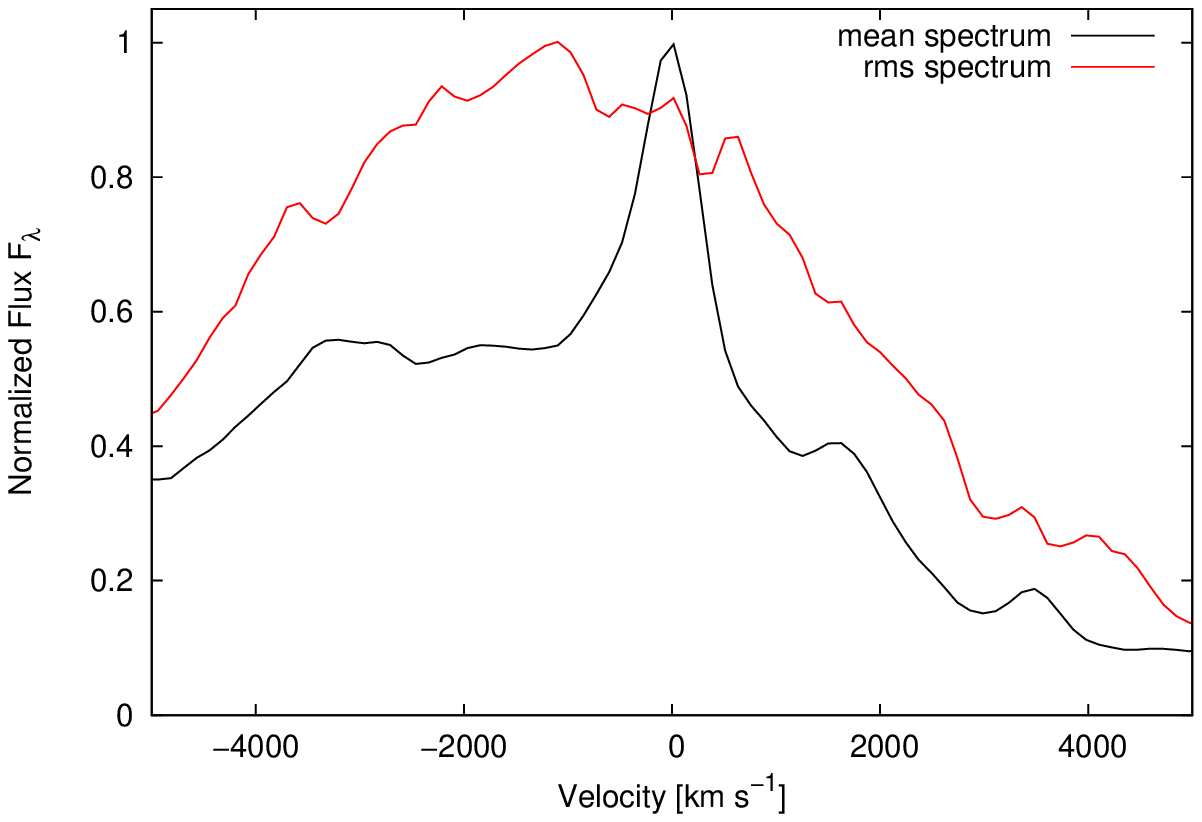}
  \caption{Normalized mean (black) and rms (red)
 line profiles of  \ion{He}{ii}\,$\lambda 4686$ in velocity space.}
  \label{hb0405meanrms.ps}
\end{figure}
\begin{figure}
 \includegraphics[width=9cm]{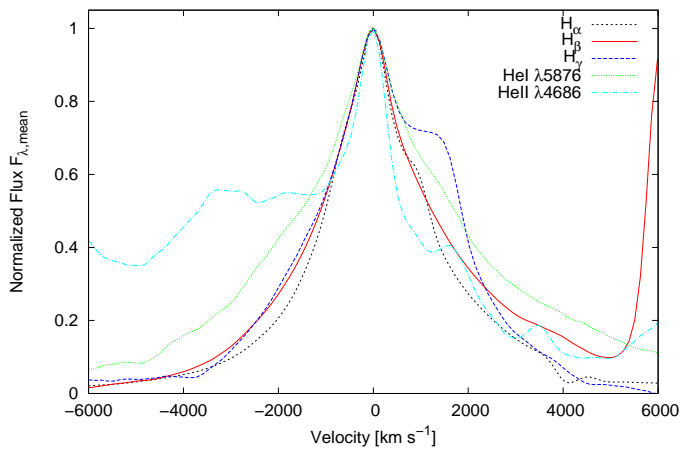}
  \caption{Normalized mean line profiles of
    H$\alpha$, H$\beta$,  H$\gamma$, \ion{He}{i}\,$\lambda 5876$,
    and \ion{He}{ii}\,$\lambda 4686$.}
  \label{hbha0405mean.ps}
\end{figure}
\begin{figure}
\vspace{-3mm}
 \includegraphics[width=9cm]{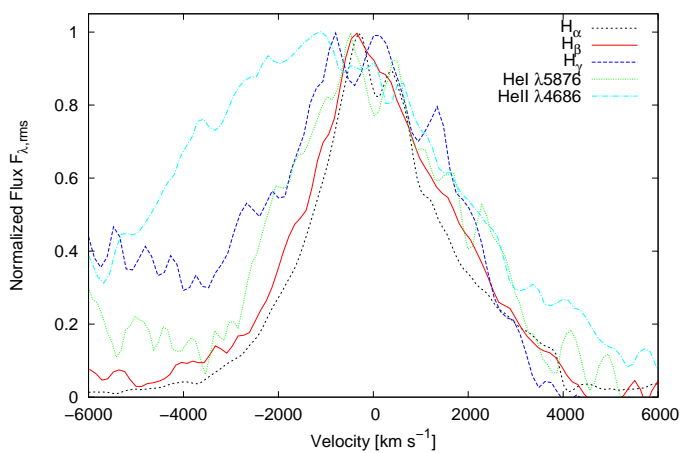}
  \caption{Normalized rms line profiles of
    H$\alpha$, H$\beta$,  H$\gamma$, \ion{He}{i}\,$\lambda 5876$,
    and \ion{He}{ii}\,$\lambda 4686$. The H$\gamma$ wing blueward of
    2000~km s$^{-1}$ is probably due to poor continuum modeling.}
  \label{hbha0405mean.ps}
\end{figure}
%
%quite  uncertain
%because of the H$\alpha$ line blending with the [SII]-lines
%and the H$\beta$ line blending with
%the [OIII]-lines, the uncertainty being as large as ~3~000  
%km s$^{-1}$. The red end of the H$\alpha$ line profile is beyond our observed
%spectral range.
%The central component and the outer rms components are not arranged
%strictly symmetrical 
%with respect to v = 0~km s$^{-1}$.
%
%
They are presented in Figs.\,6 to 12
in velocity space.
We show in Figs.\,6 to 10 the normalized mean and rms
line profiles of the individual lines. 
The rms spectra illustrate the variations in the line profile segments
during our variability campaign.
The strong blue wing of the H$\gamma$ rms line in Fig.~8 (i.e. short-wards
 of $-2000$ km s$^{-1}$)
 is probably caused
by the strong blue variability of the underlying continuum (see Fig.2).
Because there is no continuum window on
the blue side of the H$\gamma$ line in our 3C\,120 spectra we were unable
to subtract a pseudo-continuum below the line with the same precision
as for the other lines.
In Figs.\,11 and 12 we present all normalized mean profiles
of the Balmer and helium
lines and all normalized rms profiles. Thus we can compare the
different line widths FWHM and different profile shapes 
and line asymmetries.

We present in Table 7 the line widths FWHM of the mean
and rms line profiles of the Balmer and helium lines.
 Furthermore, we parameterized the line widths of the rms profiles by
their line dispersion $\sigma_{line}$ (rms widths) (Fromerth \& Melia
\citealt{fromerth00}; Peterson et al. \citealt{peterson04}).
\begin{table}
\centering
\tabcolsep+2mm
\caption{Balmer and helium line widths: FWHM of the mean
and rms line profiles and line dispersion $\sigma_{line}$ (rms width)
of the rms profiles.}
\begin{tabular}{lccc}
\hline 
\noalign{\smallskip}
Line & FWHM (mean) & FWHM (rms) & $\sigma_{line}$ (rms) \\
     & [km s$^{-1}$] & [km s$^{-1}$] & [km s$^{-1}$] \\
(1)  & (2) & (3) & (4) \\ 
\noalign{\smallskip}
\hline
\noalign{\smallskip}
H$\alpha$                        & 2205 $\pm$   74 & 2630 $\pm$  87 &  1638 $\pm$ 105  \\
H$\beta$                         & 2386 $\pm$   52 & 3252 $\pm$  67 &  1689 $\pm$  68  \\
H$\gamma$                        & 3029 $\pm$   65 & 4696 $\pm$ 979 &  1939 $\pm$  78  \\
\ion{He}{i}\,$\lambda 5876$      & 2962 $\pm$   99 & 3987 $\pm$ 128 &  2358 $\pm$ 570  \\
\ion{He}{ii}\,$\lambda 4686$     & 5821 $\pm$ 1297 & 6472 $\pm$ 132 &  3253 $\pm$ 130  \\
\noalign{\smallskip}
\hline 
\end{tabular}
\end{table}
In Table~8 we display the shifts of the line centers of the 
rms and mean line profiles. We derived the emission line centers
using only the parts of the line profiles above 75\% of the peak value.
\begin{table}
\centering
\tabcolsep+2mm
\caption{Line center (above 75\% of the peak value)
in the rms and mean line profiles.}
\begin{tabular}{lll}
\hline 
\noalign{\smallskip}
Line center & rms profile & mean profile\\
     & [km s$^{-1}$] &[km s$^{-1}$]\\
(1)  & (2) & (3) \\ 
\noalign{\smallskip}
\hline
\noalign{\smallskip}
H$\alpha$                        &$-145.5\pm 137.  $&$-19.0 \pm 18.5 $ \\
H$\beta$                         &$-191.6\pm 104.  $&$-16.7 \pm 25.1 $ \\
H$\gamma$                        &$-199.8\pm  57.  $&$+33.3 \pm 28.2 $ \\
\ion{He}{i}\,$\lambda 5876$      &$-213.4\pm 171.  $&$+11.4 \pm 19.2 $ \\
\ion{He}{ii}\,$\lambda 4686$     &$-1197.4\pm 66.  $&$-29.7 \pm 25.8 $ \\
\noalign{\smallskip}
\hline 
\end{tabular}
\end{table}

There are three clear trends in the mean and rms emission-line profiles
in 3C\,120:\\
- the individual rms profiles (FWHM) are always broader than the mean profiles.\\
- the (higher ionized) helium lines are always broader than the Balmer lines.\\
- the (higher ionized) helium rms profiles exhibit stronger blueshifts and
 asymmetries than the Balmer lines.\\
These trends indicate that the variable part of the emission-line
profiles (i.e., the rms spectrum) originates closer to the center -  where the
rotation velocity is higher - than the non-variable part.
% rms profile originates closer to the center
%than the contribution of emission line clouds the mean profile.
Furthermore, the  (higher ionized) helium lines also
originate closer to the center than the Balmer lines.
A blueshift in the rms profiles in comparison with symmetric mean profiles
can be explained by an additional outflow
component that is moving towards the observer. 
In disk-wind models the more distant receding part of the wind is
occulted by the accretion disk so that the red
side of the line profile of a high-ionization line is suppressed 
(e.g. Gaskell\citealt{gaskell09}).

\subsection{CCF analysis}
The distance of the broad-line emitting region from the central
ionizing source can be estimated in AGN by correlating
% deriving the cross-correlation function (CCF) of
the broad emission-line light curves 
with that of
 the ionizing continuum flux. A continuum light curve
in the optical is normally used as surrogate for the ionizing light curve. 
An interpolation cross-correlation function
method (ICCF) has been developed by Gaskell \& Peterson\cite{gaskell87}
to calculate the delay of the two light curves. 
We developed our own ICCF code 
(Dietrich \& Kollatschny \citealt{dietrich95}) in a similar way.
With this method we correlated the light curves
of the Balmer and helium lines of 3C\,120 with the 
continuum light curve at 5100\,\AA{}. The
cross-correlation functions ICCF($\tau$) are presented in Fig.~13.
\begin{figure}
\includegraphics[width=90mm,angle=0]{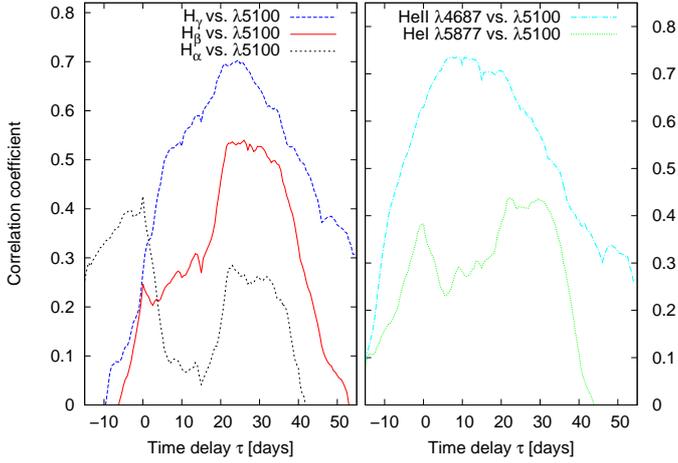}
\caption{Cross-correlation functions CCF($\tau$)
           of the Balmer and helium line light curves
           compared with the continuum light curve at 5100\,\AA{}.}
  \label{3.7_line_ccfs.eps}
\end{figure}
We derived the centroids of these ICCF, $\tau_{cent}$,
by using only the part of the CCF above 80\% of the peak value.
It has been shown by Peterson et al.\cite{peterson04} 
that a threshold value of 0.8 $r_{max}$ is generally a good choice.
We determined the uncertainties in our cross-correlation results by
calculating the cross-correlation lags many times using
a model-independent Monte Carlo method known as
\textit{flux redistribution/random subset selection} (FR/RSS).
This method has been described
by Peterson et al.\cite{peterson98}.
Here the error intervals correspond to 68\% confidence levels. 

The final results of the ICCF analysis are given in Table\,9.
\begin{table}
\centering
\tabcolsep+16mm
\caption{Cross-correlation lags of the Balmer and helium line light curves
     with respect to the 5100\,\AA\ continuum light curve.}
\begin{tabular}{lc}
\hline 
\noalign{\smallskip}
%Line & \multicolumn{1}{c}{$\tau_{cent}$} \\
Line & \multicolumn{1}{c}{$\tau$} \\
     & \multicolumn{1}{c}{[days]}\\
(1)  & \multicolumn{1}{c}{(2)}\\
\noalign{\smallskip}
\hline
\noalign{\smallskip}
H$\alpha$                    &   $28.5^{+9.0}_{-8.5}$\\[.7ex]
H$\beta$                     &   $27.9^{+7.1}_{-5.9}$\\[.7ex]
H$\gamma$                    &   $23.9^{+4.6}_{-3.9}$\\[.7ex]
\ion{He}{i}\,$\lambda 5876$  &   $26.8^{+6.7}_{-7.3}$\\[.7ex]
\ion{He}{ii}\,$\lambda 4686$ &   $12.0^{+7.5}_{-7.0}$\\[.7ex]
\noalign{\smallskip}
\hline 
\end{tabular}
\end{table}
The delay of
%Balmer line-averaged BLR size (H$\beta$ and H$\alpha$)
the integrated H$\beta$ line   
with respect to the continuum light
curve at 5100\,\AA\ 
corresponds to $28.5^{+9.0}_{-8.5}$~light-days.
The other Balmer lines and the \ion{He}{i}\,$\lambda 5876$ show similar
delays of 24 to 28 light-days. 
The delay of
the integrated \ion{He}{ii}\,$\lambda 4686$ line only   
corresponds to $12.0^{+7.5}_{-7.0}$~light-days.
It is known that there is a radial BLR stratification
in AGN (e.g. Kollatschny\citealt{kollatschny03}).  
The higher ionized lines show broader line widths (FWHM) and originate
closer to the center (Fig.\,14).
In a similar way, the variability amplitude of the
integrated emission lines is correlated with the distance
of the line-emitting region to the central ionizing source (Fig.\,15).
We present in Fig.\,14 the theoretical relation between distance and
 line width for different black hole masses 
based on the mass formula given in section 3.5. 
For this diagram we used the corrected
rotational velocities $v_{rot}$ given in Table\,10.
\begin{figure}
\includegraphics[width=60mm, angle=270] {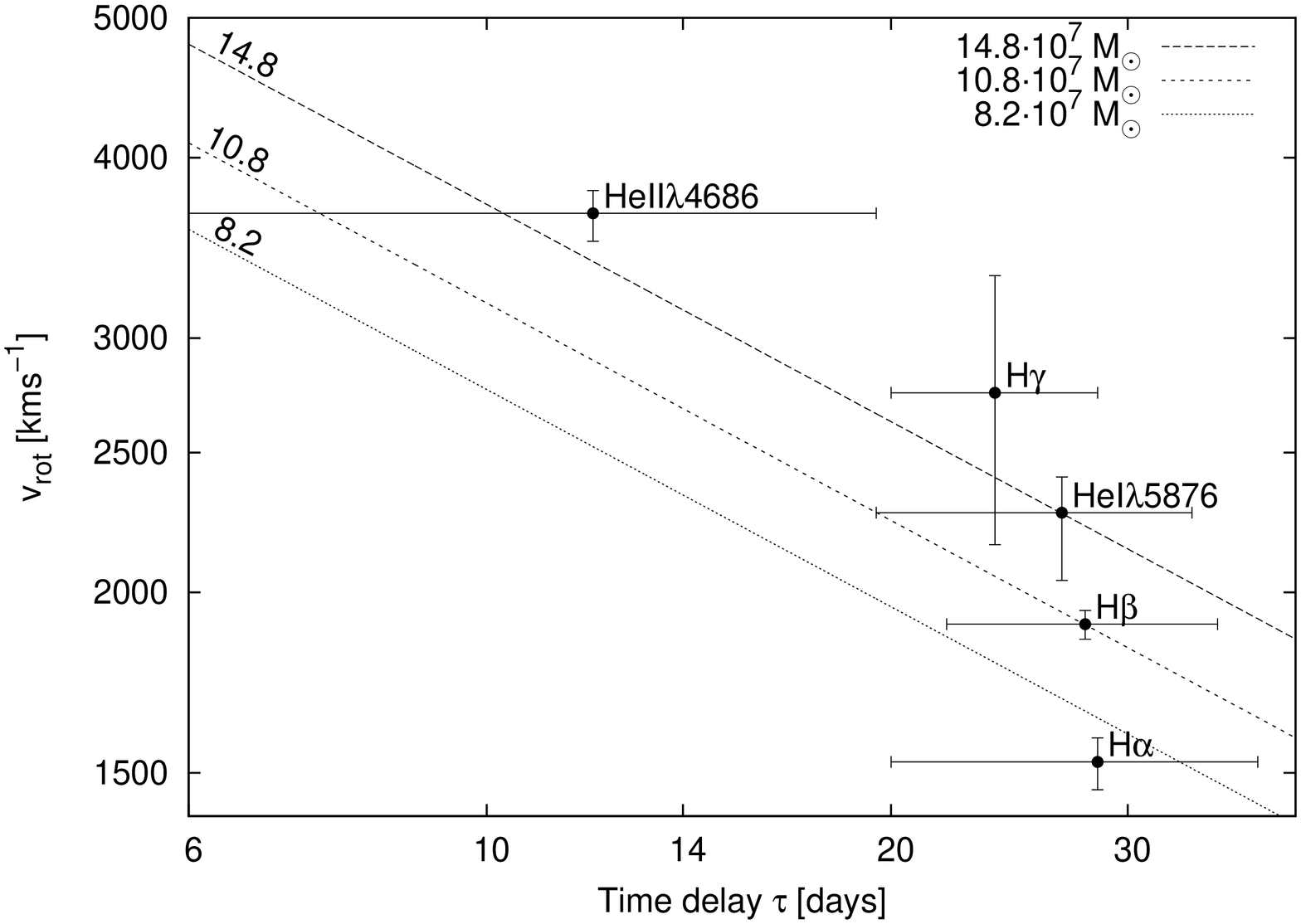}
\caption{Line width of the emission lines (FWHM) as function of their
 time delay $\tau$ (i.e.distance to
 the center). The dotted and dashed lines correspond to virial masses of
14.8, 10.8, and 8.2$\times 10^7$ M$_{\odot}$.}
  \label{alttau_fwhm.eps}
%\end{figure}
%
%\begin{figure}
%  \includegraphics[width=85mm,angle=0] {3.12_tau_vs_fmaxfmin.eps}
%  \includegraphics[width=85mm,angle=0] {fig15_rmax_tau_korelat.eps}
  \includegraphics[width=85mm,angle=0] {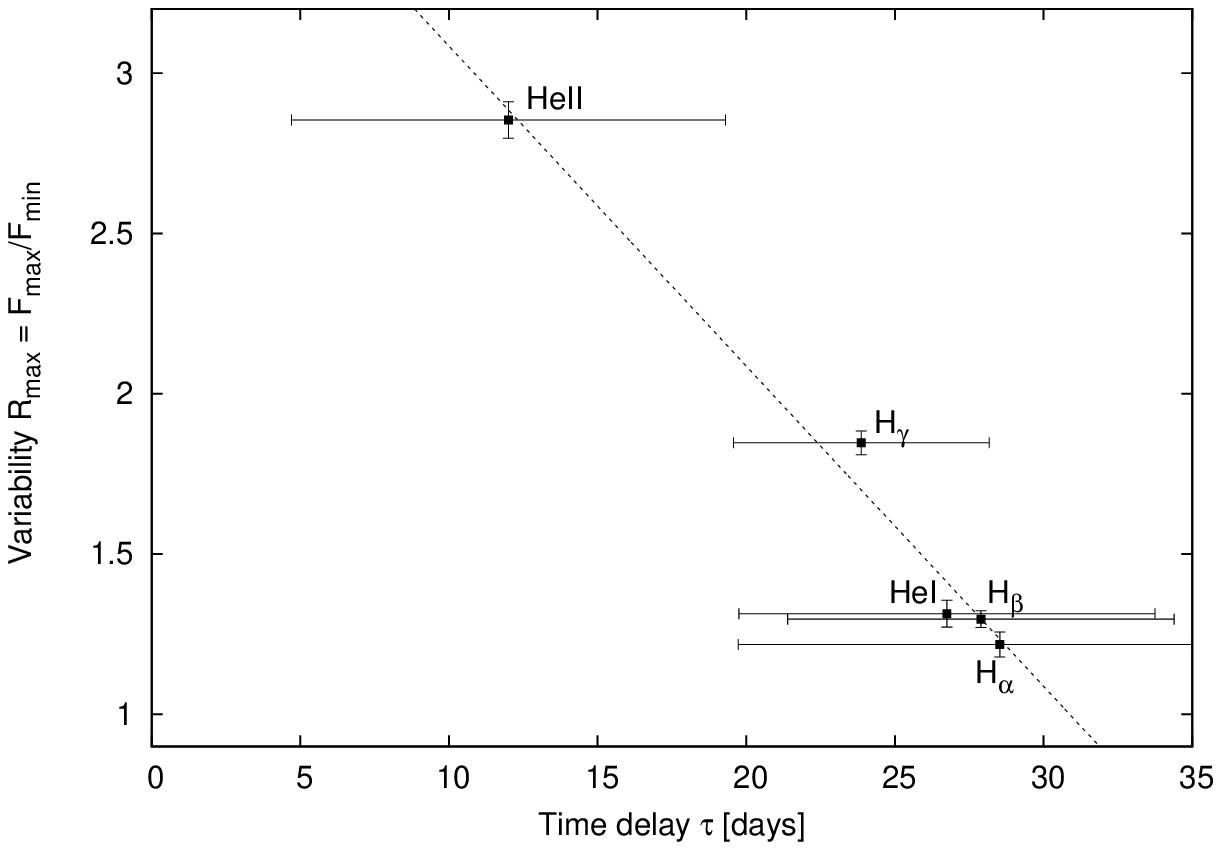}
\caption{Variability amplitude of the integrated emission lines
 as function of their time delay $\tau$ (i.e.distance to
 the center). The dotted line is the best fit to the data.}
  \label{3.12_tau_vs_fmaxfmin.eps}
\end{figure}
It shows the known trend that the \ion{He}{ii}\,$\lambda 4686$ line originates
closer to the center than the Balmer lines and the
\ion{He}{i}\,$\lambda 5876$ line. 

\subsection{Central black hole mass}

The central black hole mass in AGN can be
derived from
the width of the broad emission line profiles based on the assumption 
that the gas dynamics are dominated by the central massive object,
by evaluating
$M = f\,c\,\tau_{cent}\,\Delta\,v^{2}\, G^{-1}  $.
%$M = f\,c\,\tau_{cent}\,\sigma^{2}\, G^{-1}  $.\\
%\[ M = \frac{f c}{G} v^{2} G^{-1} R . \]
%
The characteristic distance of the line-emitting region
 $\tau_{cent}$
 is given by the centroid of the individual cross-correlation
functions of the emission-line variations
compared with the continuum variations
 (e.g. Koratkar \& Gaskell \citealt{koratkar91}; Kollatschny \& Dietrich
\citealt{kollatschny97}).
The characteristic velocity $\Delta v$ of the emission-line
region can be estimated from the FWHM of the rms profile
 or from the line dispersion $\sigma_{line}$.

The scaling factor $f$ in the equation above is on the order
of unity and depends
on the kinematics, structure, and orientation of the BLR.
This scaling factor may differ from galaxy to galaxy 
depending on whether we see the
central accretion disk including the BLR from the edge or face-on. 
We wish to compare our value of the central black hole mass in 3C\,120
with values of the black hole mass derived by other authors
(i.e., Grier et al.\citealt{grier12}) and therefore,
we adopted their mean value of f=5.5.

%\begin{table}
%\centering
%\tabcolsep+2.0mm
%\caption{FWHM
%% and line dispersion $\sigma_{line}$ 
%of the Balmer and Helium rms line profiles, time delays of their light
%curves with respect to the continuum light curve at 5100\AA{}, as well as the
%%rms widths (FWHM) of the H$\beta$ and H$\alpha$ emission lines,
%derived central black hole masses.
%}
%\begin{tabular}{lccc}
%\hline 
%\noalign{\smallskip}
%Line    &   FWHM (rms)      &   $\tau_{cent}$   &   M$_{\text{FWHM}}$           \\ 
%        &   [km s$^{-1}$]   &  [days]           &   [$10^7 M_{\odot}$]       \\
%(1)     &     (2)           &    (3)            &   (4)                        \\
%\noalign{\smallskip}
%\hline
%\noalign{\smallskip}
%H$\alpha$                     &  2630 $\pm$  87 &  $28.5^{+9.0}_{-8.5}$   &   5.1 $\pm$ 1.6       \\
%H$\beta$                      &  3252 $\pm$  67 &  $27.9^{+7.1}_{-5.9}$   &   7.7 $\pm$ 1.6       \\
%H$\gamma$                     &  4696 $\pm$ 235 &  $23.9^{+4.6}_{-3.9}$   &  13.7 $\pm$ 2.8       \\
%\ion{He}{i}\,$\lambda 5876$   &  3987 $\pm$ 128 &  $26.8^{+6.7}_{-7.3}$   &  11.1 $\pm$ 3.0       \\
%\ion{He}{ii}\,$\lambda 4686$  &  6472 $\pm$ 132 &  $12.0^{+7.5}_{-7.0}$   &  13.0 $\pm$ 8.0       \\
%\noalign{\smallskip}
%\hline 
%\end{tabular}
%\end{table}
%

Based on the derived delay of the integrated H$\beta$ line ($\tau_{cent} = 27.9\pm6.5$\,days)
and on the H$\beta$ line width (FWHM(rms)$ = 3252 \pm  67\,$km\,s$^{-1}$),
we calculate a black hole mass of 
\[ M = 3.1 \pm 0.7 \times 10^{8} M_{\odot} . \]
%
%$\pm$ basierend auch auf f=5.5 und etwas kleineren FWHM=2539/Sigma=1514 Werten
%sowie tau=25.9. 
%
However, here we did not correct for the contribution of turbulent motions
to the width of the line profiles so far. This is presented in section 3.7
in more detail.
After correcting the H$\beta$ line width (FWHM) for its
contribution of turbulent motions we derived a black hole mass of
\[ M = 10.8 \pm 2.6 \times 10^{7} M_{\odot} . \]

Grier et al.\cite{grier12} derived a black hole mass of 
$M = 6.7 \pm 0.6 \times 10^{7} M_{\odot}$ based on
the line dispersion  $\sigma_{line}$ of her H$\beta$ data.
The two values agree well. 

Based on the similarly corrected line widths (FWHM)
of the additional Balmer and
helium lines and on their derived delays,
we calculated black hole masses of
$7.7\pm 2.3\,(H\alpha)$,  
$19.4\pm 8.9\, (H\gamma)$,
$14.8 \pm 4.6\, (\ion{He}{i}\,\lambda 5876)$, and
$17.3 \pm 10.5 \times 10^{7} M_{\odot}\, (\ion{He}{ii}\,\lambda 4686)$. 
All these BH masses agree with each other within the error limits.

%For all emission lines we derive an  weighted average
%black hole mass
%\[ M = 12.0 \pm 3.5 \times 10^{7} M_{\odot}  \]
%for 3C\,120 (?? Wert noch genau berechnen).}

\subsection {2D CCF of the Balmer (H$\alpha$, H$\beta$,
H$\gamma$) and helium\,I, II line profiles}
%----------------------------------------------------------------------------- 

In this section we investigate in more detail the profile variations of
the Balmer and helium lines in 3C\,120.
We proceed in the same way as for the line profile variations in
Mrk\,110 (Kollatschny \& Bischoff \citealt{kollatschny02}; Kollatschny
\citealt{kollatschny03}) and Mrk\,926
(Kollatschny \& Zetzl\citealt{kollatschny10}).  

We sliced the velocity profiles of the Balmer and helium lines  
into velocity segments of widths $\Delta$v = 400 km s$^{-1}$.
This value of
400 km s$^{-1}$ corresponds to the spectral resolution of our observations.
Then we measured the intensities of all subsequent
velocity segments 
from $v = -3\,800$ until $+3\,800$\,km\,s$^{-1}$ 
 and compiled their light curves.
The central line segment was integrated from
$v = -200$ until $+200$\,km\,s$^{-1}$. Light curves of the central H$\beta$
segment, and of selected blue and red segments are shown in Fig. 16. 
%
%At first sight all the light curves of the line segments look similar.
%However, there are clear differences between the outer and inner segments
%as well as between the red and blue wings.
%
For comparison, the light curve of the continuum flux at 5100\,\AA\
is given as well.

We computed CCFs of all
line segment ($\Delta{}v=400$\,km\,s$^{-1}$) light curves of the Balmer and
helium lines
with the 5100\,\AA{} continuum light curve.
The derived delays of the segments compared with the 5100\,\AA{}
continuum light curve are shown in 
Figs.~17 to 21 as a function of distance to the line center.
These 2D CCFs are presented in gray scale.
The white lines in Figs.~17--21 delineate the
contour lines of the correlation coefficient
at different levels.
The black curves show computed escape velocities for central masses of
$3.5, 7, 14  \times 10^{7} M_{\odot}$ (from bottom to top, see, e.g.,
Kollatschny \& Bischoff \citealt{kollatschny02}).

The light curves of the line centers
are mostly delayed by 20 to 35 days with respect
to the continuum variations---except for HeII$\lambda$4686.
The outer line wings at distances of 2000 to 3000 km s$^{-1}$
respond much faster to continuum variations
than the inner line profile segments, by 0 to only 20 days.
The outer blue wing of  the H$\beta$ line (shortward of $-3000$\,km\,s$^{-1}$)
is blended with the red wing of the HeII$\lambda$4686 line (see Fig.2).
Therefore the response shortward of -3000 km s$^{-1}$ is influenced
by the HeII$\lambda$4686 line. 
In contrast to the Balmer and HeI lines,
the HeII$\lambda$4686 line originates at a distance
of only about 11 light days (Fig.~21, Table~9). 
And there is no indication for a longer delay of
the line center with respect to the line wings.
% in this HeII$\lambda$4686 line.
In the discussion section we compare
our observed velocity delay maps in more detail
with model calculations of echo images from the BLR
and with other
% 2-D CCF($\tau$,$v$)
observations.

%----------------------------------------------------------------------------- 
%                                                
   \begin{figure*}
    \includegraphics[width=15cm,angle=0]{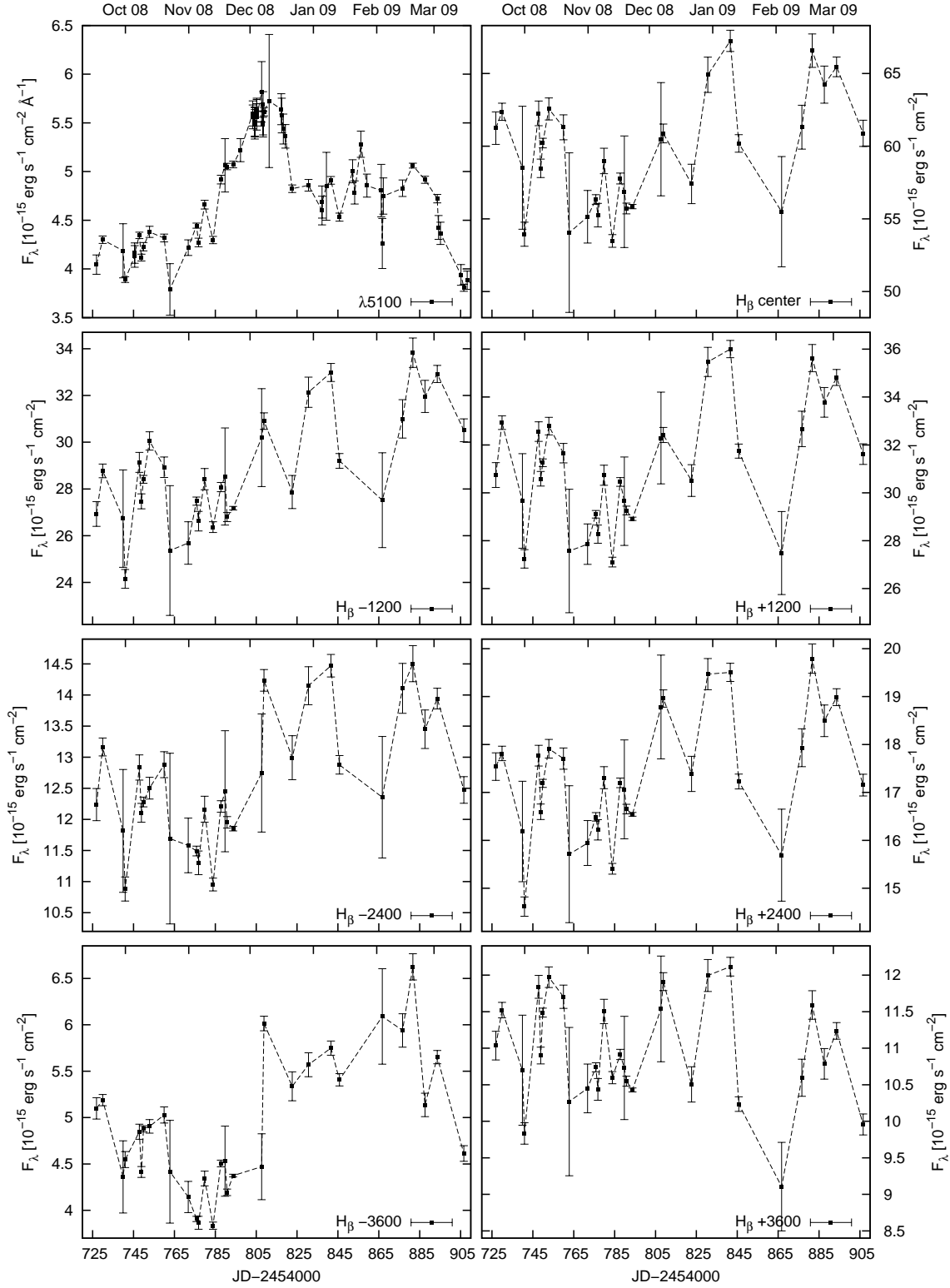}
%fig16_velohb_int.eps
      \caption{Light curves of the continuum flux at 5100\,\AA\
  and of selected H$\beta$ line segments
   (in units of  10$^{-15}$ erg s$^{-1}$ cm$^{-2}$):  H$\beta_{center}$, and segments at
v = $+/-$1~200, $+/-$2~400, $+/-$3~600 km s$^{-1}$. 
}
         \label{5.3_hb_v_lightcurves.eps}
   \end{figure*}
%
%______________________________________________________________
%\clearpage
%\newpage
%------------------------------------------------------------------------------
%
\begin{figure}
%\begin{figure*}
% \hbox{
%\includegraphics[width=83mm]{5.7_2d_ccf_ha.eps}
%\includegraphics[width=83mm]{2dccfha_v2.eps}
%\includegraphics[width=83mm]{fig17_2dccfha.eps}
\includegraphics[width=83mm]{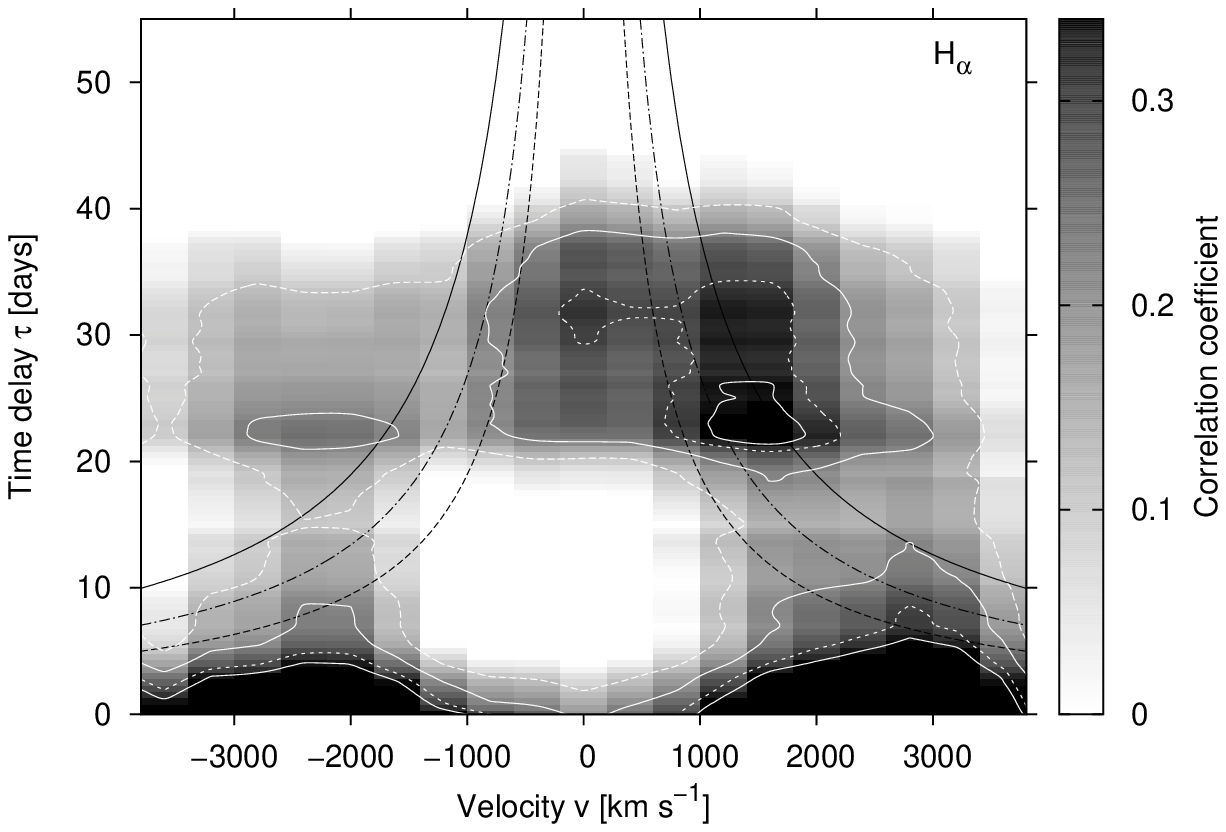}
%\includegraphics[bb=40 90 380 700,width=9.12cm,angle=270]{5.7_2d_ccf_ha.eps}
%}
%       \vspace*{-2mm} 
%       \vspace*{25mm}
  \caption{2D CCF($\tau$,$v$) showing the correlation of the H$\alpha$ line  
segment light curves with the continuum light curve
% at  5180\,\AA{}
as a function of velocity and time delay (gray scale).
Contours of the correlation coefficients are overplotted at levels of
%0.37, 0.32, 0.24, 0.17, 0.08, 0.01 (solid lines).
%0.28 / 0.2 / 0.1 (white lines Neu Dez. 13).
0.32, 0.28, 0.2, 0.1 (white lines).
The black curves show computed escape velocities for central masses of
$3.5, 7, 14  \times 10^{7} M_{\odot}$ (from bottom to top).
}
   \label{5.7_2d_ccf_ha.eps}
%\end{figure*}
%\end{figure}
%
%----------------------------------------------------------------------------- 
%------------------------------------------------------------------------------
%
%\begin{figure}
%\begin{figure*} 
%\hbox{
%\includegraphics[width=83mm]{5.8_2d_ccf_hb.eps}
%\includegraphics[width=83mm]{2dccfhb_v2.eps}
%\includegraphics[width=83mm]{fig18_2dccfhb.eps}
%\includegraphics[width=83mm]{fig18_2dccfhb2.eps}
\includegraphics[width=83mm]{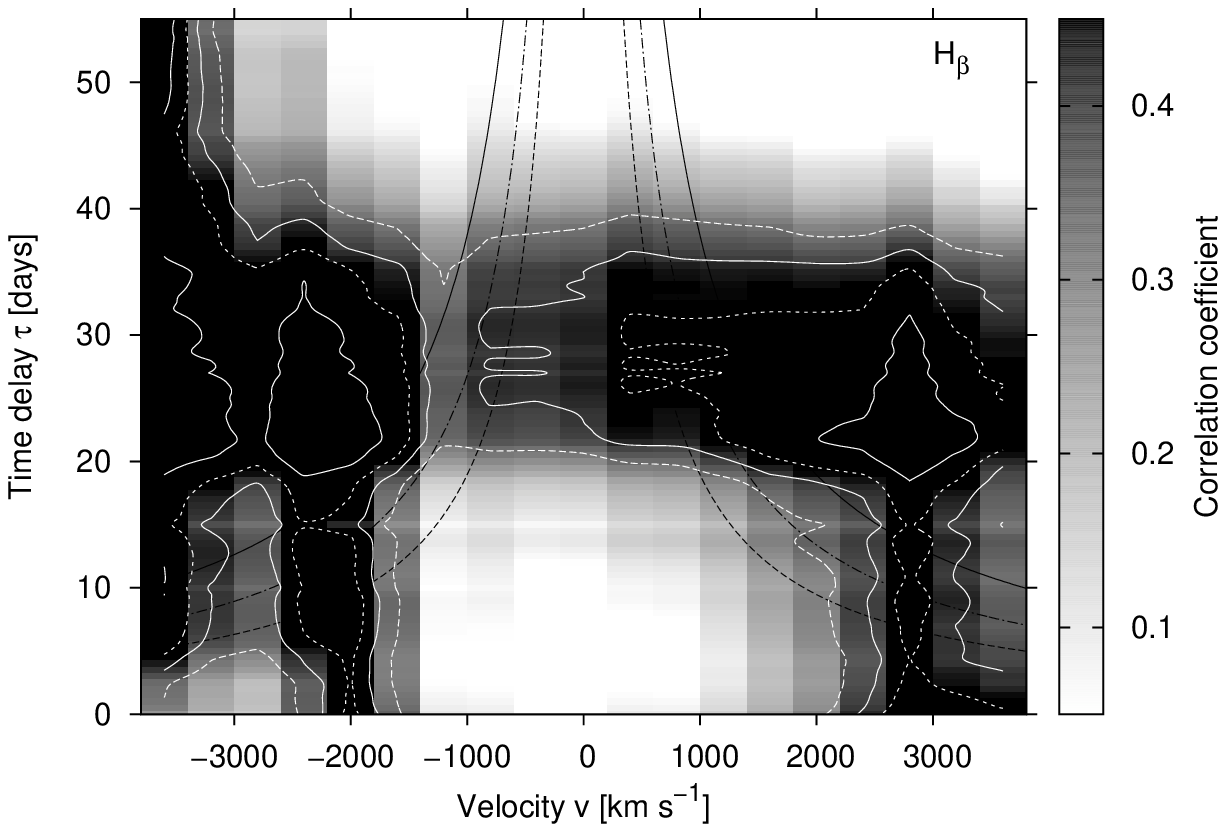}
%\includegraphics[bb=40 90 380 700,width=9.12cm,angle=270]{5.8_2d_ccf_hb.eps}
%}
%       \vspace*{-2mm} 
%       \vspace*{25mm}
  \caption{2D CCF($\tau$,$v$) showing the correlation of the H$\beta$ line  
segment light curves with the continuum light curve
% at  5180\,\AA{}
as a function of velocity and time delay (gray scale).
Contours of the correlation coefficients are overplotted at levels of
%0.37, 0.35, 0.31, 0.20, 0.12, 0.06 (solid lines).
%0.59 / 0.51 / 0.41 (white lines Neu Dez. 13).
0.53, 0.46, 0.4, 0.32 (white lines).
Black curves as in Fig.~17.
%The dashed curves show computed escape velocities for central masses of
%$3.5, 4, 14  \times 10^{7} M_{\odot}$ (from bottom to top).
}
   \label{5.8_2d_ccf_hb.eps}
%\end{figure*}
%\end{figure}
%
%------------------------------------------------------------------------------
%
%\begin{figure}
%\begin{figure*} 
%\hbox{
%\includegraphics[width=83mm]{5.9_2d_ccf_hg.eps}
%\includegraphics[width=83mm]{2dccfhg.eps}
%\includegraphics[width=83mm]{2dccfhg_v2.eps}
\includegraphics[width=83mm]{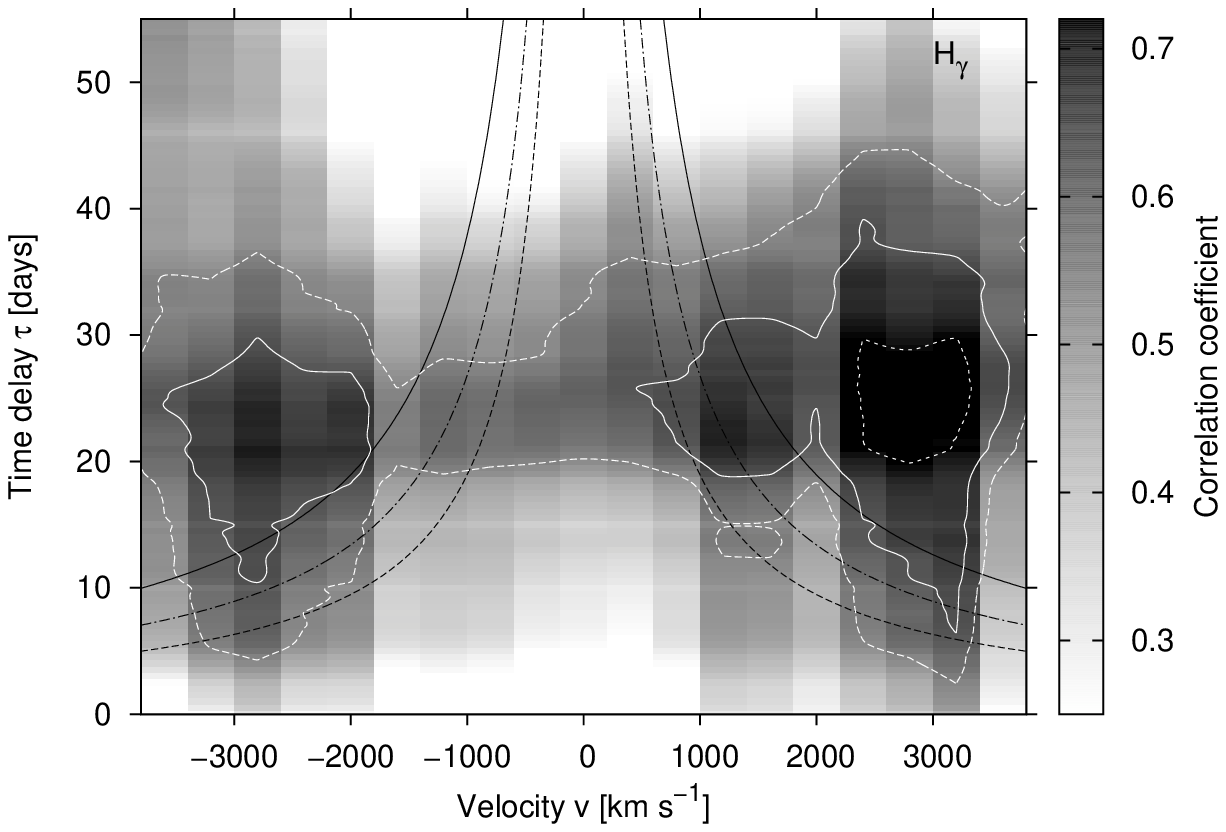}
%\includegraphics[bb=40 90 380 700,width=9.12cm,angle=270]{5.9_2d_ccf_hb.eps}
%}
%       \vspace*{-2mm} 
%       \vspace*{25mm}
  \caption{2D CCF($\tau$,$v$) showing the correlation of the H$\gamma$ line  
segment light curves with the continuum light curve
% at  5180\,\AA{}
as a function of velocity and time delay (gray scale).
Contours of the correlation coefficients are overplotted at levels of
%0.46, 0.39, 0.32, 0.25  (solid lines).
0.71, 0.62, 0.53 (white lines).
Black curves as in Fig.~17.
%The dashed curves show computed escape velocities for central masses of
%$3.5, 4, 14  \times 10^{7} M_{\odot}$ (from bottom to top).
}
   \label{5.9_2d_ccf_hg.eps}
%\end{figure*}
\end{figure}
%------------------------------------------------------------------------------
%
%------------------------------------------------------------------------------
%
\begin{figure}
%\begin{figure*} 
%\hbox{
%\includegraphics[width=83mm]{5.10_2d_ccf_hei.eps}
%\includegraphics[width=83mm]{2dccfhei_v2.eps}
\includegraphics[width=83mm]{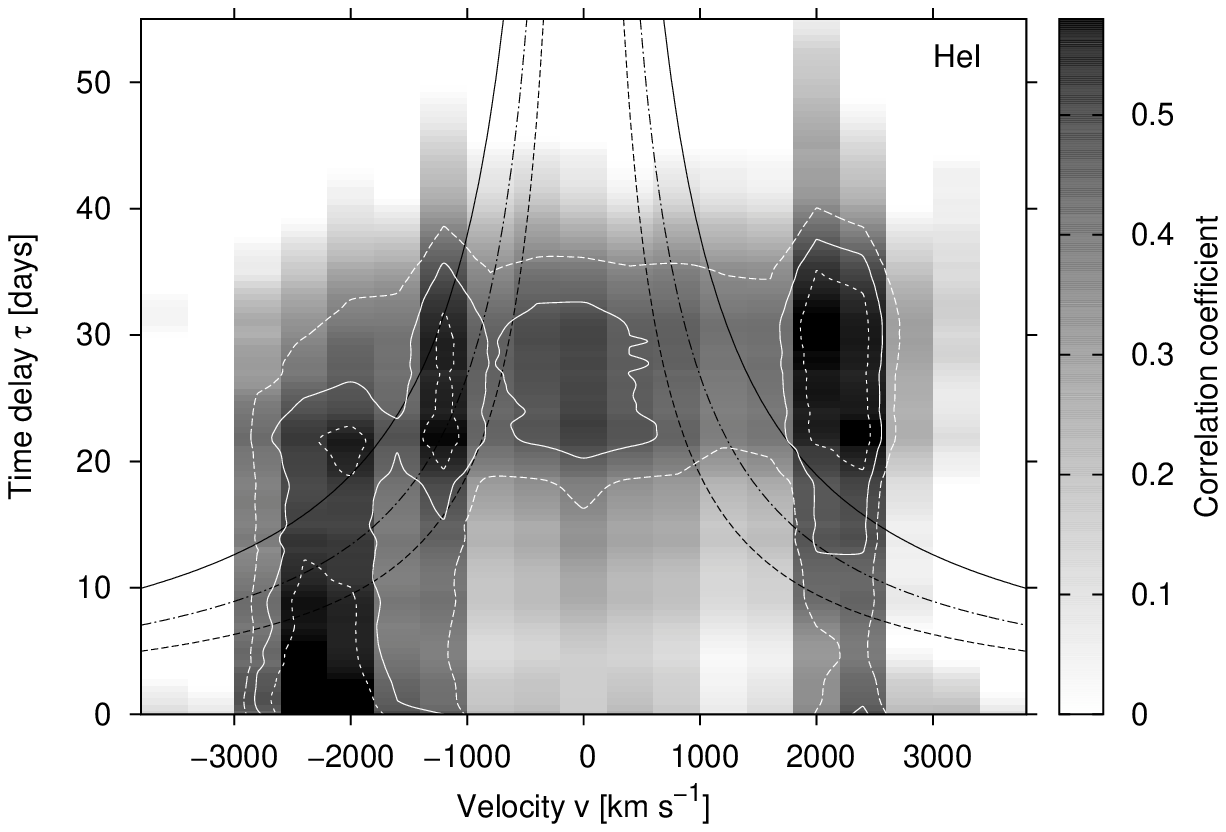}
%\includegraphics[bb=40 90 380 700,width=9.12cm,angle=270]{5.10_2d_ccf_hei.eps}
%}
%       \vspace*{-2mm} 
%       \vspace*{25mm}
  \caption{2D CCF($\tau$,$v$) showing the correlation of the  
\ion{He}{i}\,$\lambda 5876$ line  
segment light curves with the continuum light curve
% at  5180\,\AA{}
as a function of velocity and time delay (gray scale).
Contours of the correlation coefficients are overplotted at levels of
%0.54, 0.49, 0.42, 0.32, 0.24  (solid lines).
0.52, 0.44, 0.34 (white lines).
Black curves as in Fig.~17.
%The dashed curves show computed escape velocities for central masses of
%$3.5, 4, 14  \times 10^{7} M_{\odot}$ (from bottom to top).
}
   \label{5.10_2d_ccf_hei.eps}
%\end{figure*}
%\end{figure}
%
%------------------------------------------------------------------------------
%
%\begin{figure}
%\begin{figure*} 
%\hbox{
%\includegraphics[width=83mm]{2dccfheii2_neu.eps}
%\includegraphics[width=83mm]{Fig21_2dccfheii2_engl.eps}
%\includegraphics[width=83mm]{2dccfheii_v2.eps}
%\includegraphics[width=83mm]{fig21_2dccfheii.eps}
\includegraphics[width=83mm]{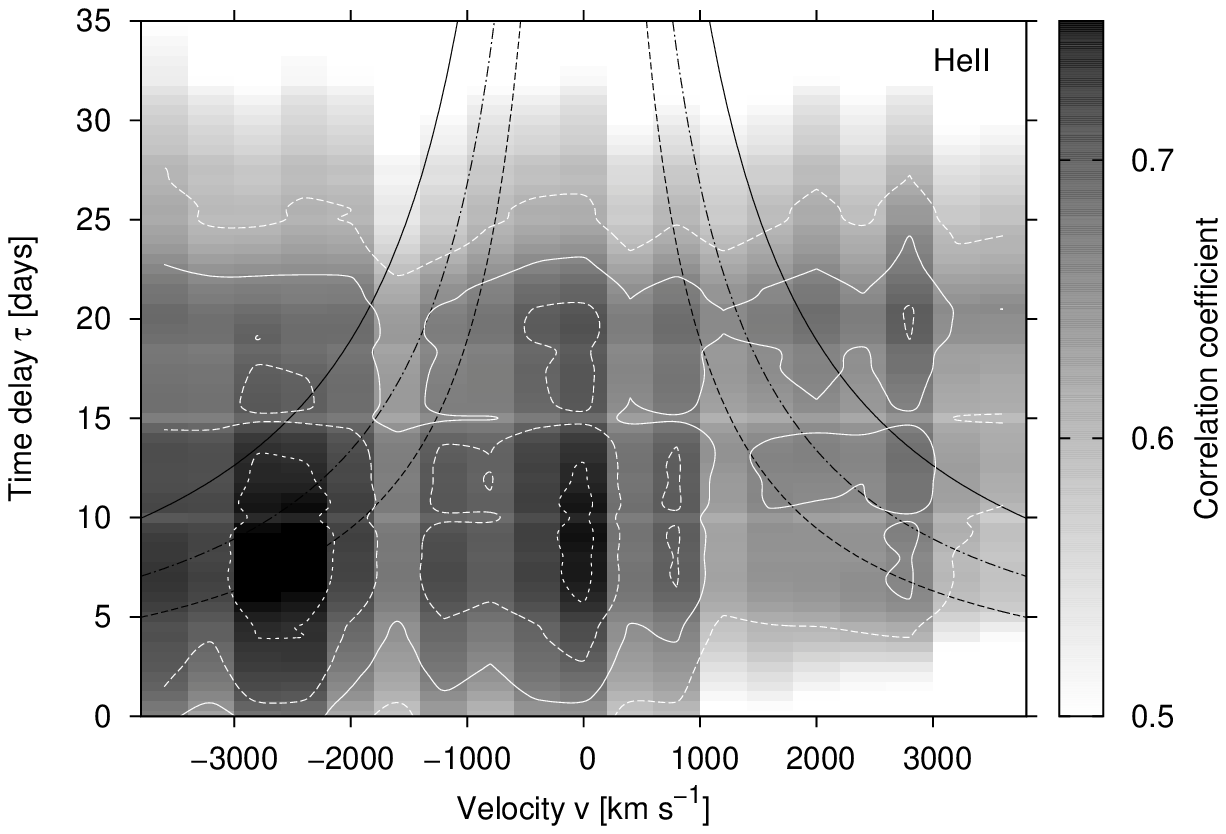}
%\includegraphics[bb=40 90 380 700,width=9.12cm,angle=270]{5.8_2d_ccf_hb.eps}
%}
%       \vspace*{-2mm} 
%       \vspace*{25mm}
  \caption{2D CCF($\tau$,$v$) showing the correlation of the  
\ion{He}{ii}\,$\lambda 4686$ line
segment light curves with the continuum light curve
% at  5180\,\AA{}
as a function of velocity and time delay (gray scale).
Contours of the correlation coefficients are overplotted at levels of
%0.66, 0.62, 0.57, 0.51 (solid lines).
%0.75 / 0.69 / 0.63 (white lines Neu Dez. 13).
%0.75 / 0.69 / 0.65 / 0.62.
0.72, 0.68, 0.64, 0.59 (white lines). 
Black curves as in Fig.~17.
%The dashed curves show computed escape velocities for central masses of
%$3.5, 4, 14  \times 10^{7} M_{\odot}$ (from bottom to top).
}
   \label{5.11_2d_ccf_heii.eps} 
%\end{figure*}
\end{figure}
%
%
%\newpage

%\clearpage

\subsection {Vertical BLR structure in 3C~120}

 We demonstrated in recent papers (Kollatschny
 \& Zetzl\citealt{kollatschny11,kollatschny13a,kollatschny13b,kollatschny13c})
that we are able to make statements about the BLR structure in AGN
based on variability
studies in combination with line profile studies.
The broad emission-line profiles in AGN can be
parameterized by the ratio of
their FWHM to their line dispersion
$\sigma_{\mathrm{line}}$.
There exists a general relation between the full-width at half maximum and
 the line-width ratio 
FWHM/$\sigma_{\mathrm{line}}$ for the individual emission lines.
\begin{table*}%[htbp]
    \centering
       \leavevmode
       \tabcolsep2mm 
        \newcolumntype{d}{D{.}{.}{-2}} 
        \newcolumntype{p}{D{+}{\,\pm\,}{-1}}
        \newcolumntype{K}{D{,}{}{-2}}
\caption{Line profile parameters and radius and height of
 the line-emitting regions
for individual emission lines in 3C\,120. The H$\beta$ parameters
and the continuum luminosities are given for three observing campaigns.
%($M_{\mathrm{BH}} = 79.6\times10^{6}\,  \mathrm{M}_{\odot}$, $R_{\mathrm{S}}
% = 9.08\times10^{-3}\,\mathrm{ld} = 2.35\times10^{12}\,\mathrm{cm}$)
}
\begin{tabular}{lKKKKKlKd|c}
 \htopline
\hspace{3mm} Line & \mcr{FWHM/$\sigma$} &\mcc{$v_{\text{turb}}$}&\mcc{$v_{\text{rot}}$} &\mcc{Radius}
                  & \mcc{Height} & \mcr{$H/R$} & \mcc{Height$_{\text{corr}}$} & \mcc{$H_{\text{corr}}/R$}&\mcc{$\log10(\lambda{}L_{\lambda})$}\\
\hspace{3mm}      &&\mcc{[\kms{}]}&\mcc{[\kms{}]} &\mcc{[ld]}
                  & \mcc{[ld]} & \mcr{} & \mcc{[ld]} & \mcr{}&\mcr{[erg\,s$^{-1}$]}\\
\hmidline   
\Nl{He}{ii}{4686} &1.99,\pm{0.09}&1993,^{+347}_{-319} &3661,^{+135}_{-160}&12.0,^{+7.5}_{-7.0}&6.5,\pm{4.2} &0.54&3.0,\pm{2.2} &0.25&$44.12\pm{0.07}$\\
\Hg               &2.42,\pm{0.16}&486,^{+132}_{-128}  &2750,^{+564}_{-592}&23.9,^{+4.6}_{-3.9}&4.2,\pm{1.4} &0.18&3.7,\pm{1.4} &0.15&$\cdot{}$\\
\Nl{He}{i}{5876}  &1.69,\pm{0.27}&1139,^{+609}_{-483} &2271,^{+133}_{-232}&26.8,^{+6.7}_{-7.3}&13.4,\pm{8.2}&0.50&9.4,\pm{7.7} &0.35&$\cdot{}$\\
\Hb               &1.93,\pm{0.09}&507,^{+73}_{-70}    &1901,^{+42}_{-45}  &27.9,^{+7.1}_{-5.9}&7.4,\pm{2.2} &0.27&5.9,\pm{1.8} &0.21&$\cdot{}$\\
\Ha               &1.61,\pm{0.12}&548,^{+108}_{-101}  &1526,^{+60}_{-66}  &28.5,^{+9.0}_{-8.5}&10.2,\pm{3.8}&0.36&13.1,\pm{4.6}&0.46&$\cdot{}$\\
\hmidline
\Hb{} (p04)       &1.89,\pm{0.18}&243,^{+48}_{-49} &1291,^{+109}_{-110}&38.1,^{+21.3}_{-15.3}&7.2,\pm{4.3}&0.19&11.8,\pm{6.8}&0.31&$44.01\pm{0.05}$\\
\hmidline
\Hb{} (g12)       &1.68,\pm{0.32}&453,^{+106}_{-110} &1480,^{+280}_{-294}&25.6,^{+2.4}_{-2.4}&7.8,\pm{2.6}&0.30&6.9,\pm{2.4}&0.27&$43.87\pm{0.05}$\\
\hbotline  
\end{tabular}
\end{table*}

The line width FWHM reflects the line broadening due to
rotational motions of the broad-line gas of the intrinsic Lorentzian
profiles that are associated with
turbulent motions. Different emission lines are connected with
different turbulent velocities. 

Here we model the observed line width ratios
 FWHM/ $\sigma_{\mathrm{line}}$ versus the line width FWHM of 3C\,120
(Fig.~22, Table~10) in the same way as for other Seyfert galaxies
(Kollatschny
 \& Zetzl\citealt{kollatschny11,kollatschny13a,kollatschny13b,kollatschny13c}).
Based on their observed line widths (FWHM) and line-width ratios
FWHM/$\sigma_{\mathrm{line}}$, we plot in Fig.\,22 the locations of all
observed emission lines in 3C\,120. We derived the rotational velocities
that belong to the individual lines from their positions between
 the vertical dashed lines that represent different $v_{\text{rot}}$.
The given H$\beta$ line width ratios in Fig.~22 are based on three
different variability campaigns: the variability campaign presented in this
paper and two additional variability campaigns carried out
by Peterson et al.\cite{peterson04} and Grier et al.\cite{grier12}. 
They are marked with {\it p04} and {\it g12} in Figs.~22--26
and Table~10.

%
%------------------------------------------------------------------------------
%
\begin{figure}[t]
%\begin{figure*} 
%\hbox{
\includegraphics[width=62mm, angle=270]{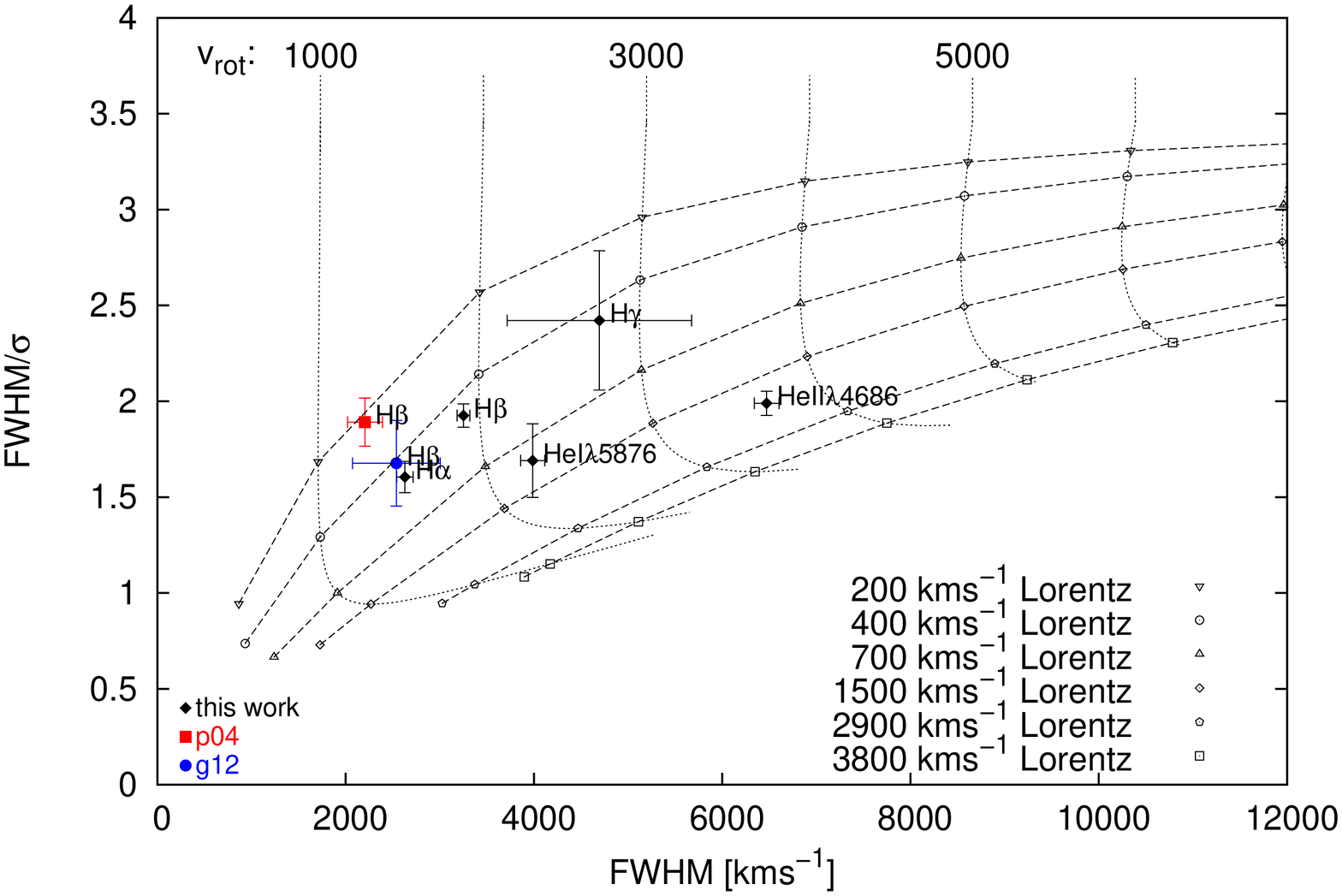}
%\includegraphics[bb=40 90 380 700,width=9.12cm,angle=270]{5.8_2d_ccf_hb.eps}
%}
%       \vspace*{-2mm} 
%       \vspace*{25mm}
  \caption{Observed and modeled line-width ratios
 FWHM/ $\sigma_{\mathrm{line}}$ versus line width FWHM in 3C\,120.
 The dashed curves represent the corresponding theoretical
line-width ratios based on rotational line-broadened
 Lorentzian profiles (FWHM = 200 to
3800\,km\,s$^{-1}$). The rotation velocities reach
from 1000 to 6000\,km\,s$^{-1}$
(curved dotted lines from left to right).}
   \label{fwhmsigma_3c120.ps} 
%\end{figure*}
\end{figure}
%
%------------------------------------------------------------------------------
 
We determined the heights of the line-emitting regions above the midplane 
on the basis of the turbulent velocities that belong to the individual
emission lines as for other Seyfert galaxies.
We used the following mean turbulent velocities:
425 km\,s$^{-1}$ for H$\gamma$,
400 km\,s$^{-1}$ for H$\beta$,  
700 km\,s$^{-1}$ for H$\alpha$,
900 km\,s$^{-1}$ for \ion{He}{ii}\,$\lambda 4686$.
In Table 10 we present the derived heights above the midplane of
the line-emitting regions in 3C\,120 (in units of light-days) and the
ratio $H/R$ for the individual emission lines.
The ratio of the turbulent velocity $v_{turb}$ compared with the
rotational velocity $v_{rot}$ in the line-emitting region
gives us information on the ratio of the accretion disk
height  $H$ with respect to the accretion disk radius $R$
 of the line-emitting regions
as presented in Kollatschny \& Zetzl \cite{kollatschny11,kollatschny13a} :
\begin{equation}
\label{eq:HtoR}
  H/R = (1/\alpha) (v_{turb}/v_{rot}). 
\end{equation}
The unknown viscosity parameter $\alpha$ is assumed to be constant
and to have a value of one.
We have not yet derived the mean turbulent velocity connected to the
\ion{He}{i}\,$\lambda 5876$ emission-line region in AGN in our earlier papers.
Based on Fig.~22, we computed a value on the order of 800 km\,s$^{-1}$
for \ion{He}{i}\,$\lambda 5876$ from
this single variability campaign.
%As all the turbulent velocities for the different emission lines
%in this variability campaign are relatively small we assume a slightly
%higher 'general' turbulent velocity of 800 km\,s$^{-1}$ for
%the \ion{He}{i}\,$\lambda 5876$ line---i.e., between the Balmer lines 
%and the \ion{He}{ii}\,$\lambda 4686$.
%The following results are independent of this assumption.

We show in Fig.~23 the BLR structure of  3C\,120 as a function of
distance to the center and height above the midplane. 
The H$\beta$ emission regions observed at different epochs are connected
by a solid line. The errors are large for the H$\beta$ region of 
the variability campaign of the years
1989 -- 1996 (Peterson et al.\citealp{peterson98})
because of poor sampling.
The dot at radius zero gives the size of the Schwarzschild radius
$R_{\mathrm{S}}
 = 1.2\times10^{-2}\,\mathrm{ld} = 3.2\times10^{13}\,\mathrm{cm}$ for a
% = 2.97\times10^{-3}\,\mathrm{ld} = 7.67\times10^{11}\,\mathrm{cm}$ for a
black hole mass
(with $M=10.8\times10^{7}M_{\sun}$)
% according to Table 10)
% (with $M=2.6\times10^{7}M_{\sun}$)
 multiplied by a factor of twenty.
The label at the top of the figure gives
the distances of the line-emitting regions
in units of the Schwarzschild radius.
%
%------------------------------------------------------------------------------
% 
%\vspace*{15mm} 
\begin{figure}[t]
%\begin{minipage}[t]{0.475\textwidth}
\includegraphics[width=7.6cm,angle=0]{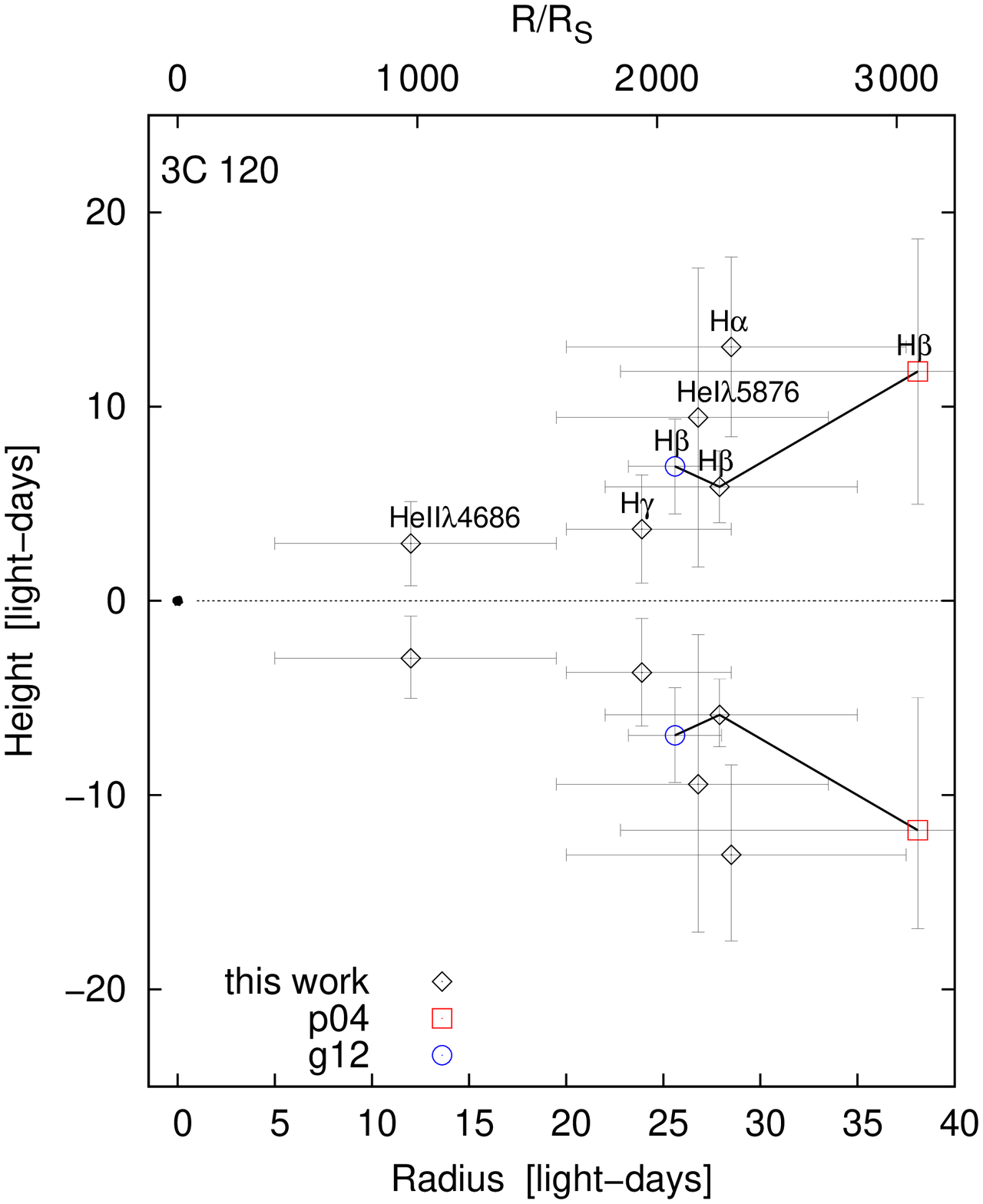} 
  \caption{Structure of the BLR in 3C\,120.
   The H$\beta$  positions for the three variability campaigns
   are connected by a solid line.
   The dot at radius zero has the size of a Schwarzschild black hole
 {\bf (for $M_{BH}=10.8\times10^{7}M_{\sun}$)}
 multiplied by a factor of twenty.}
   \label{disc_3c120.ps}
%\end{minipage}
%\hfill
%\begin{minipage}[t]{0.475\textwidth}
%\includegraphics[width=7.6cm,angle=0]{disc_ngc3783.ps} 
%  \caption{NGC~3783 BLR structure. Same as Fig. 10, but
%   based on corrected turbulent velocities $v_{\mathrm{turb}}$.} 
%   \label{disc_ngc3783corr.eps}
%\end{minipage}
\end{figure}
%
%----------------------------------------------------------------------------- 
%
The \ion{He}{ii}\,$\lambda 4686$ line originates
at the shortest distance from the center and smallest height above the midplane
in comparison with the Balmer and HeI lines,
as has been seen before in other galaxies.
H$\alpha$ originates at a larger distance from the midplane than H$\beta$.
This has also been seen before in NGC~7469 (Kollatschny \& Zetzl
\citealt{kollatschny13c}). Here we see the same effect in 3C~120.

\section{Discussion}

We monitored 3C\,120 in the years 2008 and 2009. At this time the
AGN was in a brighter
state (Fig. 3) than in observations taken earlier during the years
1989 -- 1996 (Peterson et al.\citealt{peterson98}) or later during the year
2010 (Grier et al.\citealt{grier12})
(see the mean continuum luminosities given in Table~10).
In our discussion section we highlight the line-profile variations
in 3C\,120 during our campaign in comparison with other variability campaigns.
Furthermore, we discuss the results regarding the BLR structure in
3C\,120 in comparison with the BLR structure in other AGN.  
% At these epochs the continuum
% flux at 5100\,\AA{} was of the order of 3--6 $\times$
%10$^{-15}$\,erg\,s$^{-1}$\,cm$^{-2}$\,\AA$^{-1}$ and 3--4 $\times$
%10$^{-15}$\,erg\,s$^{-1}$\,cm$^{-2}$\,\AA$^{-1}$ respectively.

\subsection{Structure and kinematics in the BLR of 3C~120}

The response of emission-line segments compared with the variable
ionizing continuum does not only give information about the distance of
the line-emitting regions, but also on their kinematics
in comparison with model calculations. 
These 2D CCF($\tau$,$v$) are
mathematically very similar to 2D response functions $\Psi$ (Welsh
 \citealt{welsh01}).

Our 2D cross-correlation functions CCF($\tau$,$v$) of
the Balmer (H$\alpha$, H$\beta$, H$\gamma$) and helium 
(\ion{He}{i}\,$\lambda 5876$, \ion{He}{ii}\,$\lambda 4686$) line
segment light curves with the continuum light curve at  5100\,\AA{}
are presented in Figs.~17 to 21
as a function of velocity and time delay (gray scale).
There is a general trend to be seen in the Balmer and 
\ion{He}{i}\,$\lambda 5876$ lines that
the emission-line wings at distances of 2\,000 to 3\,000 km s$^{-1}$ 
from the line center respond much faster than the central region.
The centers of these lines respond with a delay of 25 -- 30 light-days
with respect to the ionizing continuum,
while the response in the wings is much faster, with delays of 0 -- 20
light-days.

Grier et al.\cite{grier13} monitored the variability in the
H$\beta$, H$\gamma$, and \ion{He}{ii}\,$\lambda 4686$ lines of 3C~120
one and a half years later than we did. 
They also reported a lack of prompt response in the
line centers of H$\beta$ and H$\gamma$ in 3C~120.
The faster response in the Balmer line wings
compared with that of to the line center
has been seen in variability campaigns of other AGN as well,
for example, in Mrk\,110 (Kollatschny \citealt{kollatschny03}),
NGC\,5548 (Kollatschny \& Dietrich \citealt{kollatschny96} and
Denney et al. \citealt{denney09}), SBS~1116  (Bentz et al. \citealt{bentz09}),
NGC\,4593 (Kollatschny \& Dietrich \citealt{kollatschny97}). 
The faster response of the line wings is explained
by accretion disk models for the line-emitting regions
(e.g. Perez et al.\citealt{perez92}, or Welsh \& Horne\citealt{welsh91}).
The BLR Keplerian disk model of
Welsh \& Horne \cite{welsh91} (their Fig.~1c) 
agrees remarkably well with our Balmer and
HeI$\lambda$5876 line observations.

Many of the Seyfert galaxies that have been monitored spectroscopically
show indications for additional velocity components in the
velocity-delay maps, that is, a stronger and faster response in the red
or the blue wing.
An earlier response in the blue line wing than in the red wing is
connected with outflow motions in the models of  
Perez et al.\cite{perez92} and Horne et al.\cite{horne04}.
In disk-wind models the blue side of a line profile responds to changes
in the ionizing continuum with almost no lag, while the red side of the line
follows with twice the lag of the line as a whole
(Gaskell \citealt{gaskell09}).
An earlier response in the red wings is connected with inflow motions.
On the other hand,
an earlier response of the red line wing than in the blue line wing
is predicted in  the spherical wind and
 disk-wind models of 
%(K\"{o}nigl \& Kartje \cite{koenigl94},
Chiang \& Murray \cite{chiang96}.
%Blandford \& Begelman \cite{blandford99}).
In their models the line-emitting gas shows a radial outward
velocity component in addition to the rotation. 
%Furthermore, the influence of optical thickness effects is unknown.
%It might be that
%the highest velocity gas seen in the wings of the profiles
%is (at least in some cases) optically thin  and does not correspond to
%the continuum variations (Denney et al. \citealt{denney10}).

Only a few galaxies present
a shorter response in the blue H$\beta$ wings than in the red wings
in the 2D cross-correlation functions
as seen in Mrk~817 or NGC~3227 (Denney et al.\citealt{denney10}).
On the other hand,
most of the monitored Seyfert galaxies show a shorter and stronger
response in their red Balmer wings than in the blue wings.
Typical examples are Mrk\,110 (Kollatschny\citealt{kollatschny03}),
Arp~151 (Bentz et al.\citealt{bentz10}), Mrk~1501, and
PG~2130 (Grier et al.\citealt{grier13}).
Grier et al.\cite{grier13} found a stronger response in the red wings
in both the  H$\beta$ and \ion{He}{ii}\,$\lambda 4686$ emission-line
profiles of 3C~120 as well. They explained their observing result
by infall in addition to rotational motions in the BLR.   

We monitored 3C~120 one and a half year earlier than Grier et al.\cite{grier13}
at a time when this galaxy was in a bright state (see Table~10).
The log of the mean continuum luminosity amounted to
$\log10(\lambda{}L_{\lambda})$ = 44.12
compared with 43.87 one and a half years later
(Grier et al.\citealt{grier13}).
The pattern of our H$\beta$ velocity delay map is similar to that of
Grier et al.\cite{grier13}, demonstrating a stronger response in the red wing.
The pattern of the H$\alpha$ line we observed during our campaign
shows a similar response as well. However, our 
\ion{He}{i}\,$\lambda 5876$ and \ion{He}{ii}\,$\lambda 4686$ lines
exhibit a stronger response in the blue wings than in
the red wings.
Furthermore, the blue wing in the \ion{He}{ii}\,$\lambda 4686$ line shows a
shorter response than the red wing.
This is the opposite to what was observed by
Grier et al.\cite{grier13} one and a half years
later when the galaxy was in a lower state.
%
%Irgendwo noch schreiben, dass wir fuer den Leuchtkraftvergleich
%mit Grier in Table 9 einen Korrekturfaktor von 1.2152 angebracht haben,
%da Grier einen absoluten OIII Fluss
%von 3.67$\times$10$^{-13}$\,erg\,s$^{-1}$\,cm$^{-2}$ angenommen hat.
%
A stronger and shorter response in the blue line wings 
is attributed to outflow motions
in the models of Horne et al.\cite{horne04}, for example.  
This points to outflow motions
% in the models of e.g. Horne et al.\cite{horne04}  
when the galaxy 3C~120  was in a higher activity state.
This applies particularly to
the \ion{He}{ii}\,$\lambda 4686$ line, which
originates closer to the central ionizing source.
Outflow  motions are in accordance with the evidence for
stronger variability in the blue line wings
and the blue line asymmetries based on our rms profiles. 
Radial velocity offsets due to mass outflows in AGN have been discussed 
before, for instance by  Crenshaw et
al.\cite{crenshaw10} and references therein. 

There are indications in a few other galaxies
that the response in the line wings
varied with time, for example, the response in the \ion{C}{iv}\,$\lambda 1550$
line in NGC~5548 (Kollatschny \& Dietrich \citealt{kollatschny96}).
Another interpretation for the varying response might be
off-axis variability (Gaskell \& Goosmann \citealt{gaskell13}).

\subsection{Vertical BLR structure in a sample of AGN}

We deduced the broad-line region geometry of 3C\,120 in section 3.7
(see Fig.~23). 
Now we compare the spatial distribution of the line emitting region
in 3C\,120 with those
in other galaxies:
NGC~7469, NGC~3783,  NGC~5548, and 3C~390.3
(Kollatschny \& Zetzl\citealt{kollatschny13c}).
We present in Fig.~24 the spatial distribution of the
line-emitting regions in 3C\,120 (based on observations at three epochs)
compared with
NGC~5548 (two epochs for the highly ionized lines, 13 epochs for  H$\beta$)
and compared with NGC~7469, NGC~3783, and 3C~390.3
as a function of distance to the center and height above the midplane.
The two axes scales in Fig.~24 are linear in units of light-days.
We assumed that the accretion disk structures are arranged
symmetrically to the midplane. 

%------------------------------------------------------------------------------
% 
\vspace*{5mm} 
\begin{figure}[t]
\includegraphics[width=8.0cm,angle=0]{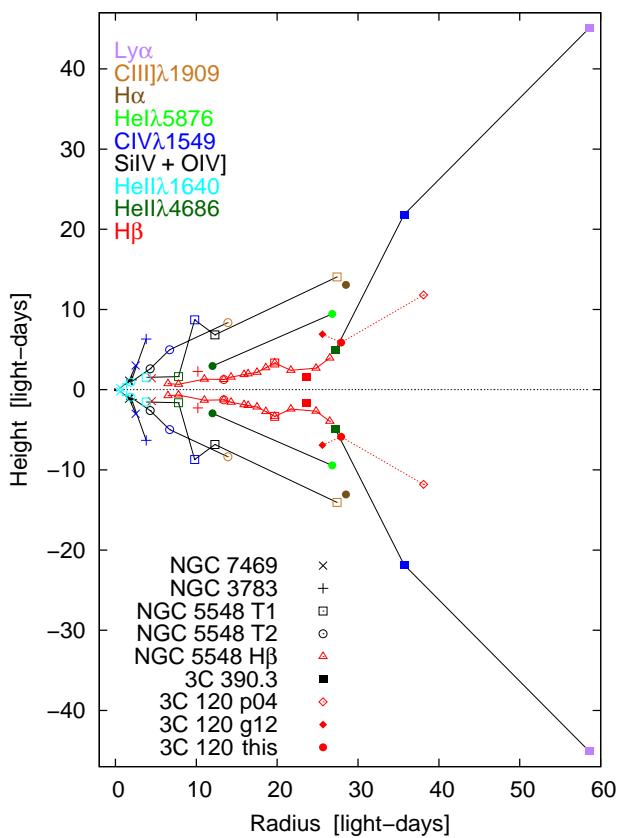} 
%\includegraphics[width=8.0cm,angle=0]{disc_all_ld.ps} 
%       \vspace*{5mm} 
%       \vspace*{-2mm} 
  \caption{BLR structures in 3C\,120 (for three epochs), NGC~7469, NGC~3783, 
NGC~5548 (for two and 13 epochs), and 3C~390.3
as a function of distance to the center and height above the midplane
(based on corrected turbulent velocities $v_{\mathrm{turb}}$).
The highly ionized (non-Balmer) lines of the individual galaxies are connected
 by a solid line. The H$\beta$ emitting regions are drawn in red.
 The three H$\beta$ line-emitting regions of  3C\,120  are connected
 by a dot-dashed red line and those 
of NGC~5548 (13 epochs) are connected by a solid red line.
% of 20 Schwarzschild radii.
}
   \label{disc_all_ld.ps}
\end{figure}
%
%----------------------------------------------------------------------------- 
%

Different emission lines are highlighted by various colors
and the different galaxies are marked by various symbols.
The H$\beta$ emitting regions are drawn in red.
The three H$\beta$ line-emitting regions of  3C\,120 
based on three different variability campaigns are connected
 by a dot-dashed red line and those 
of NGC~5548 (13 epochs) are connected by a solid red line.
3C\,120 is the most luminous AGN.
Therefore their radial H$\beta$ extension is largest for 3C\,120
as expected (e.g. Kaspi et al.\citealt{kaspi05}). 
Different emission lines originate in different regions and at different
distances from the center.
The highly ionized (non-Balmer) lines of the individual galaxies are
also connected
by a solid line to illustrate general trends in the BLR structures
of the different galaxies (see Kollatschny \& Zetzl\citealt{kollatschny13c}).
The H$\beta$ emitting regions always originate
%at lower height-to-radius ratios (up to a factor of ten)
%and they originate
closer to the midplane than the high-ionization lines.
The line widths (with respect to the individual lines)
control the height of the line-emitting 
regions above the midplane.
NGC~7469 and 3C~120, for example, show the
H$\beta$ profiles with the narrowest line widths.
They originate at the largest distances above the midplane compared with the
other AGN.
Therefore, the broad emission-line regions are not simply scaled-up versions 
that only depend on the central luminosity
(and central black hole mass).

To highlight this we show in
Fig.~25 the height-to-radius ratio for the H$\beta$ line emitting regions 
in 3C~120, NGC~7469, NGC~3783, NGC~5548,
and 3C~390.3 as a function of the H$\beta$ line width.
This plot is based on the observed $H/R$ values
(Tab.~10 and Tab.~1 in 
Kollatschny \& Zetzl\citealt{kollatschny13b},\citealt{kollatschny13c}).
% $H_{\text{obs}}/R$ values
%
% uncorrected turbulent velocities 
%$v_{\mathrm{turb}}$. These turbulent velocities 
%($v_{\mathrm{turb}}= 454^{+86}_{-81}$ this campaign, $243^{+48}_{-49}$ p04,
% $453^{+106}_{-110}$ g12) (soll ich das so schreiben)
%
%of 0.21 (this campaign), 0.31 (p04), and 0.27 (g12) given in Table~10. 
%
%
%\begin{figure*}
\begin{figure}
%\hbox{
%\includegraphics[width=6.0cm,angle=-90]{korr_h_corr_fwhm.ps} 
%\includegraphics[width=6.0cm,angle=-90]{korr_h_corr_r_fwhm.ps} 
%}\hbox{
%\includegraphics[width=6.0cm,angle=-90]{korr_h_obs_fwhm.ps} 
\includegraphics[width=6.0cm,angle=-90]{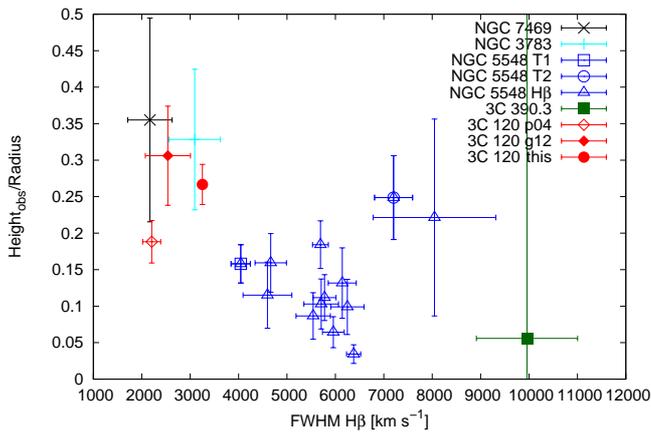} 
%}
  \caption{Height-to-radius ratio for the H$\beta$ line-emitting regions 
of 3C~120 (red), NGC~7469 (black), NGC~3783 (cyan), NGC~5548 (blue),
and 3C~390.3 (green).
}
   \label{korr_}
%\end{figure*}
\end{figure}
%
%
%(alt:  Additional
%contributions of outflowing wind components to the line widths would reduce
%the contribution of the turbulent velocities to the line profiles and
%thus reduce the numerical values of the geometrical heights.)\\
%
%
The height-to-radius ratio for H$\beta$ is highest for galaxies with
narrow emission lines and lowest for galaxies with broad lines.
The overall picture we derived for the BLR region structure
before in Kollatschny \& Zetzl\cite{kollatschny13c} is confirmed 
by the additional emission-line data of 3C~120.

\section{Summary}

We carried out
a spectroscopic monitoring campaign of the 
Seyfert~1 galaxy 3C\,120  with the 9.2m HET, and in addition an accompanying
photometric campaign with the WISE observatory in the years 2008 and 2009.
The main results of our study can be summarized as follows:

1. The broad-line region is stratified
compared  with the line widths (FWHM)
of the rms profiles, the variability amplitude of the
emission lines, 
and the distance of the line-emitting regions
from the center: The \ion{He}{ii}\,$\lambda 4686$ line originates closest
to the center, the H$\alpha$ line originates farthest away. 
Based on these data - and without correcting for the contribution of turbulent
velocity to the line profile - we derived a central black hole mass of
%\[ M = 8.5 \pm 2.1 \times 10^{7} M_{\odot} . \]
$M = 10.8 \pm 2.6 \times 10^{7} M_{\odot}$.
Within the errors this is consistent with the black hole mass derived by
Grier et al.\cite{grier12}. 

2. The velocity-delay maps of the H$\alpha$ and H$\beta$ lines
show a similar pattern as observations of H$\beta$ made by
Grier et al.\cite{grier13} one and a half years
later. The emission line wings at distances of 2\,000 to 3\,000 km s$^{-1}$ 
from the line center respond much faster 
than the central region.
The faster response of the line wings is explained
by accretion disk models.
In addition, these lines show a stronger response in their
red wings. However, the velocity-delay maps of the \ion{He}{i}\,$\lambda 5876$
and \ion{He}{ii}\,$\lambda 4686$ lines show a stronger response in the blue
wing.
Furthermore, the \ion{He}{ii}\,$\lambda 4686$ line responds
faster in the blue wing in contradiction to
the observations made one and a half
years later when the galaxy was in a lower state.
The faster response in the blue wing is an indication
for central outflow motions when this
galaxy was in a bright state during our observations.

% The 2-D cross-correlation functions CCF($\tau$,$v$) of H$\beta$ and
%H$\alpha$ outflow, inflow (changing with central luminosity)

3. The derived vertical BLR structure in 3C~120 coincides with that
of other AGN. The general trend is confirmed:
 the emission lines of narrow-line AGN originate at larger
distances from the midplane than AGN with broader
emission lines.

\begin{acknowledgements}
This work has been supported by the DFG grants Ko 857/32-1 and HA 3555/12-1,
and the Niedersachsen - Israel Research Cooperation Program ZN2318.

\end{acknowledgements}

%

%\begin{landscape}
%\setcounter{table}{3}
%\end{landscape}
\end{document}